\documentclass[amsmath,amssymb,amsfonts,apsref]{revtex4}

\usepackage{graphicx}
\usepackage{dcolumn}
\usepackage{bm}

\newcommand{\bd}{
\begin{document}}
\newcommand{\ed}{\end{document}}
\newcommand{\beq}{\begin{equation}}
\newcommand{\eeq}{\end{equation}}
\newcommand{\bef}{\begin{figure}}
\newcommand{\enf}{\end{figure}}
\newcommand{\bea}{\begin{eqnarray}}
\newcommand{\eea}{\end{eqnarray}}
\newcommand{\baR}{\begin{array}}
\newcommand{\eaR}{\end{array}}
\newcommand{\bc}{\begin{center}}
\newcommand{\ec}{\end{center}}
\newcommand{\ben}{\begin{enumerate}}
\newcommand{\een}{\end{enumerate}}
\newcommand{\bit}{\begin{itemize}}
\newcommand{\eit}{\end{itemize}}
\newcommand{\su}{\section}
\newcommand{\ssu}{\subsection}
\newcommand{\sssu}{\subsubsection}
\newcommand{\nid}{\noindent}
\newcommand{\nnb}{\nonumber}

\newcommand\cJ{{\cal J}}
\newcommand\be{{\bf e}}
\newcommand\bx{{\bf x}}
\newcommand\bu{{\bf u}}
\newcommand{\bdelta}{\mbox{\boldmath $\delta$}}
\newcommand\bthe{\mbox{{\boldmath $\theta$}}}
\newcommand\bxi{\bm{\xi}}
\newcommand\beps{{\mathbf \epsilon}}
\newcommand\bF{{\bf F}}
\newcommand\bG{{\bf G}}
\newcommand\tbG{{\tilde{\bf G}}}
\newcommand\bX{{\bf X}}
\newcommand\tu{\tilde{u}}
\newcommand\tbu{\tilde{\bu}}
\newcommand\cB{{\cal B}}
\newcommand\cC{{\cal C}}
\newcommand\cX{{\cal X}}
\newcommand\hC{\hat C}
\newcommand\hchi{\hat\chi}
\newcommand\hkappa{\hat\kappa}
\newcommand{\deq}{\stackrel {\rm def}{=}}

\bd

\title{Transmitting a signal by amplitude modulation in a chaotic network.}
\author{B. Cessac}
\affiliation{Institut Non Lin\'eaire de Nice, 1361 Route des Lucioles, 06560 Valbonne, France}
\author{J.A. Sepulchre}
\affiliation{Institut Non Lin\'eaire de Nice, 1361 Route des Lucioles, 06560 Valbonne, France}

\begin{abstract}
We discuss the ability of a network with non linear relays and chaotic dynamics
to transmit signals, on the basis of a linear response theory developed by Ruelle \cite{Ruelle} for dissipative systems.
We show in particular how the dynamics interfere with the graph topology
to produce an effective transmission network, whose topology depends on the signal, and cannot be directly
read on the ``wired'' network. This leads  one to reconsider notions such as ``hubs''.
Then, we show examples where, with a suitable choice of the carrier frequency (resonance),
one can transmit a signal from a node to another one by amplitude modulation, \textit{in spite of chaos}.
Also, we give an example where a signal, transmitted to any node via different
paths, can only be recovered by a couple of \textit{specific} nodes. This opens the possibility
for encoding data in a way  such that the recovery of the signal requires the knowledge of the carrier frequency \textit{and}
can be performed only at some specific node. 
\end{abstract}

\maketitle 

\pagebreak

{\bf
Common sense contrasts  chaotic dynamics to coherent dynamics.
For example, considering a network of dynamical units 
 having a collective chaotic dynamics,  it looks unlikely that an efficient transmission of a coherent signal  
between two arbitrary units of this  system would be feasible, since the ``butterfly effect''  would scramble and wash out  any  coherent
 signal that could be injected in the system.  
On the other hand,  there is also a common intuition that coherence can emerge through appropriate averaging procedures.   This is 
one of  the cornerstones of statistical physics which succeeds in explaining some   ``simple''  behavior of macroscopic systems 
emerging from the ``complex'' behavior of its microscopic constituents.  

Recently, new advances have shown  that some fundamental results in non equilibrium
statistical mechanics, such as  linear response theory, may be extended and generalized to dissipative 
chaotic systems \cite{Ruelle}.
 ``Dissipative'' means here that the  long term dynamics occurs on an attractor in phase space, 
a feature which is absent  in  conservative physical systems. In a previous paper \cite{CS} we have 
shown that this theory  can be applied to analyze  networks of dynamical units 
 having a chaotic and dissipative dynamics. In our case, the ``dissipation''  
comes from saturation effects in the transfer function of the units.
  This feature is known to be widespread  in biological  networks and also occurs in communication networks.
  The main result of our first study was to show the existence 
of some resonances, called ``stable'' resonances, predicted in \cite{Ruelle},  and proper to dissipative chaotic dynamics.

In the present paper we show that such dynamical networks can behave as a selective input/output system. 
  We demonstrate  that a 
network of randomly interconnected units exhibiting collective chaos, combined with a ``natural'' averaging procedure,  
can be used as an open system in which a weak coherent signal can be sent from one node  and recovered  by another receiver node.  
It is shown that such transmission  possesses a selectivity, in frequency and in the receiver node, due to the aforementioned resonances. 
These 
propagation phenomena are illustrated on a particular example, but they are  theoretically predicted 
in the firm framework of the linear response theory.  Therefore  our findings are not specific to this example.
Moreover, the method introduced here, relying on the computation and the analysis of a complex susceptibility,
can be easily applied to other systems. Indeed, stable resonances cannot
be found by studying correlations functions. Consequently, the approach proposed here goes beyond the simple correlation analysis
and brings relevant additional informations.  

Besides the possible use of our results in natural or in artificial  networks, our conclusions  point out that the study of ``networks'', 
which has recently  flourished, should not only focus on the  topology of the graph associated with  these networks, but also on its dynamics.  
If an interconnected network is composed of dynamical units,  analyzing the collective \textit{emergent} dynamics is essential to assess  the notion of 
effective ``hub'', or of effective ``connectivity'' in order  to characterise the  ability of this system  to transfer information.
}

\pagebreak

The use of statistical approaches, and especially, statistical mechanics, has been quite successful
in many research fields, sometimes far away from physics. A recent prominent example is the study
of ``networks''. There is indeed a widespread activity in this domain and many beautiful and unexpected
results have for example been derived about the topological properties of random graphs, scale free
graphs, small world, etc.. \cite{Barabasi1,Barabasi2,Newman,Blanchard1}. 
Certainly, these studies are useful to understand and anticipate the behavior
of communication networks such as Internet or mobile
phones, social networks, transport networks, epidemic propagation on networks \cite{Blanchard2}, etc...     

However, one can raise two objections to these approaches. First, a
 large part of these studies focuses on
topological properties  of the underlying graph and the nodes, whatever they represent,
are considered as ``neutral'' entities. However, in many cases, the nodes are
\textit{active} nodes behaving in a {\it nonlinear} way. For example, in a
communication network a relay regenerates (amplifies) weak
signals, but it has a finite capacity and saturates if too many
signals arrive simultaneously. A neuron has a nonlinear response
to an input current, a gene expression is determined by a nonlinear
function of the regulatory proteins concentration, etc.. These nonlinearities might modify
the network capabilities in a drastic way. This  suggests that the mere study of the
graph topological structure of a network with nonlinear nodes
is not sufficient to capture the full dynamical behavior.

The second objection concerns the statistical approach itself.
Indeed, statistical results do not deal with individuals characteristics.
But  one might be interested in knowing the characteristics of \textit{the} network he is currently
studying or using. For example,  somebody using his mobile phone is not so much interested in the 
general properties of the class of general graphs including mobile phones networks.
Rather, he is interested in   the transmission ability of the network
that \textit{he is currently using}, with its particularities.
In particular, it is not guaranteed that all nodes of \textit{this network} behave in the same way, especially
if they interact nonlinearly,
with possible excitation/inhibition effects and asymmetric interactions.
 Actually, the study
of biological networks such as neural networks or genetic networks suggests exactly
the opposite. The emerging dynamics may induce a differentiation in the effective
role of each node in the global dynamics, and the nodes are usually not interchangeable.
What happens if one perturbs  one or another node in \textit{this} network ? Are the effects the same ?
How well is  a ``signal'' sent by a given node received by another one? One may expect
that the answer depends on the pair sender/receiver. Statistical results\footnote{Note also
that to obtain analytical results one usually needs to perform a ``thermodynamic limit''
that may, in some cases, render the dynamics trivial\cite{JP}}, will wash out the particularities 
of each network realisation.\\

A careful investigation of these questions requires therefore
a proper analysis of the dynamical system corresponding to the network currently used and
may require the development of new tools and methods. 
In some situations, statistical mechanics still provides
useful insight. But one has to ``adapt'' it to the present situation and
to forget somewhat the conventional  wisdom. Thus, in a recent
paper \cite{CS}, we have shown that the linear response theory, developed by Ruelle
\cite{Ruelle} for dissipative dynamical systems, can be efficiently used to characterize
the behavior of a network where the nodes have nonlinear transfer functions. That paper focused essentially
on the existence of a new type of resonances, called stable resonances, predicted by Ruelle
but never observed before. These resonances only exist in systems where the volume is dynamically
contracted in the phase space and they are not observed in the power spectrum. In our case, 
this contraction came from saturation effects in the transfer functions.

 In the present paper, we focus on potential applications relying on the existence of such resonances.
We are mainly interested in the ability of such a network  with a \textit{chaotic} dynamics 
to propagate a signal. On the basis of the linear response theory, we first conjecture, in section \ref{SGen},
 some   non-intuitive and unexpected properties. We then exhibit an example supporting these conjectures
in the section \ref{SignProp}.
First, we compute the Fourier transform of the linear response, called the complex susceptibility,
and show how the presence of stable resonances may violate the conventional non equilibrium
statistical mechanics wisdom about linear response, fluctuation-dissipation theorem, and relaxation to equilibrium 
(sections \ref{Ssus},\ref{Ssusvscor}).
Many of these ideas are already in Ruelle's work, but their interpretation in the context of nonlinear networks
is new. We then argue that it is indeed possible to transmit and recover a signal
in a chaotic system.  We show, in particular, how the dynamics interfere with the graph topology
to produce an effective transmission network, whose topology depends on the signal and cannot be directly
read on the ``wired'' network. This leads one to reconsider notions such as ``hubs'' (section \ref{SHub}).
Then, we show examples where, with a suitable choice of the resonance frequency,
one can transmit a signal from a node to another one by amplitude modulation, \textit{in spite of the presence of chaos} (section \ref{Smodamp}).
In addition, we give an example where a signal, transmitted to any node via different
paths, can be performed only at some specific node.  Finally, we discuss the effect of increasing the signal amplitude, going 
beyond linear response in section \ref{SNL}.

\su{General setting and purposes.}\label{SGen}

In this section we recall the main results established in \cite{CS} and state the questions
addressed in the present paper. Consider a set of $N$ nodes (or relays) connected on a graph. The link
from $j$ to $i$ is denoted by $J_{ij}$. Links are oriented ($J_{ij} \neq J_{ji}$) and signed. The sign 
mimics excitation/inhibition effects. These  effects are obviously present in biological network
but they can also exist in communication networks. For example, regulation
systems exist, designed to optimize the bandwidth capacity. These systems can balance the activity of
a given relay with another one, resulting in  an effective excitation/inhibition. Note that, in our model,
the $J_{ij}$'s do not depend on the state of the node, but the linear response theory accommodates this generalisation.
 A zero link means that there is no connection from $j$ to $i$.
In the sequel the $J_{ij}$'s are fixed and do not evolve in time.
Moreover, as argued in the introduction, we are interested in the behavior of a network having
a \textit{fixed set} of $J_{ij}$'s and we do not consider 
statistical results relying on averages
over some probability distribution for the $J_{ij}$'s. 

The activity of a node $i$ is characterized by a continuous variable $x_i$. It is determined by the
set of signals coming from the nodes connected to $i$, the signal coming from $j$
being weighted by $J_{ij}$. Denote by $u_i(t)$ the total input received  by $i$ at time $t$. This is a function
of the $x_j$'s and of the $J_{ij}$'s.  We assume that the activity
of $i$ evolves according to $x_i(t+1)=f(u_i(t))$, where $f(x)$ is a \textit{sigmoidal} transfer function with a slope $g$ (e.g.
$f(x)=tanh(gx)$). Moreover, we consider the case where $g >> 1$, namely the sigmoid is strongly nonlinear.
Note that sigmoid transfer functions are encountered in neural networks (when the neuron activity is described in
terms of frequency rates), in genetic networks (Hille function), and they may also be suitably represent
amplification/saturation effects in communication protocols such as TCP/IP. This amplification/saturation
effect is actually the most important characteristics in what follows.

\begin{figure}[!ht]
\includegraphics[height=5cm,width=5cm,clip=false]{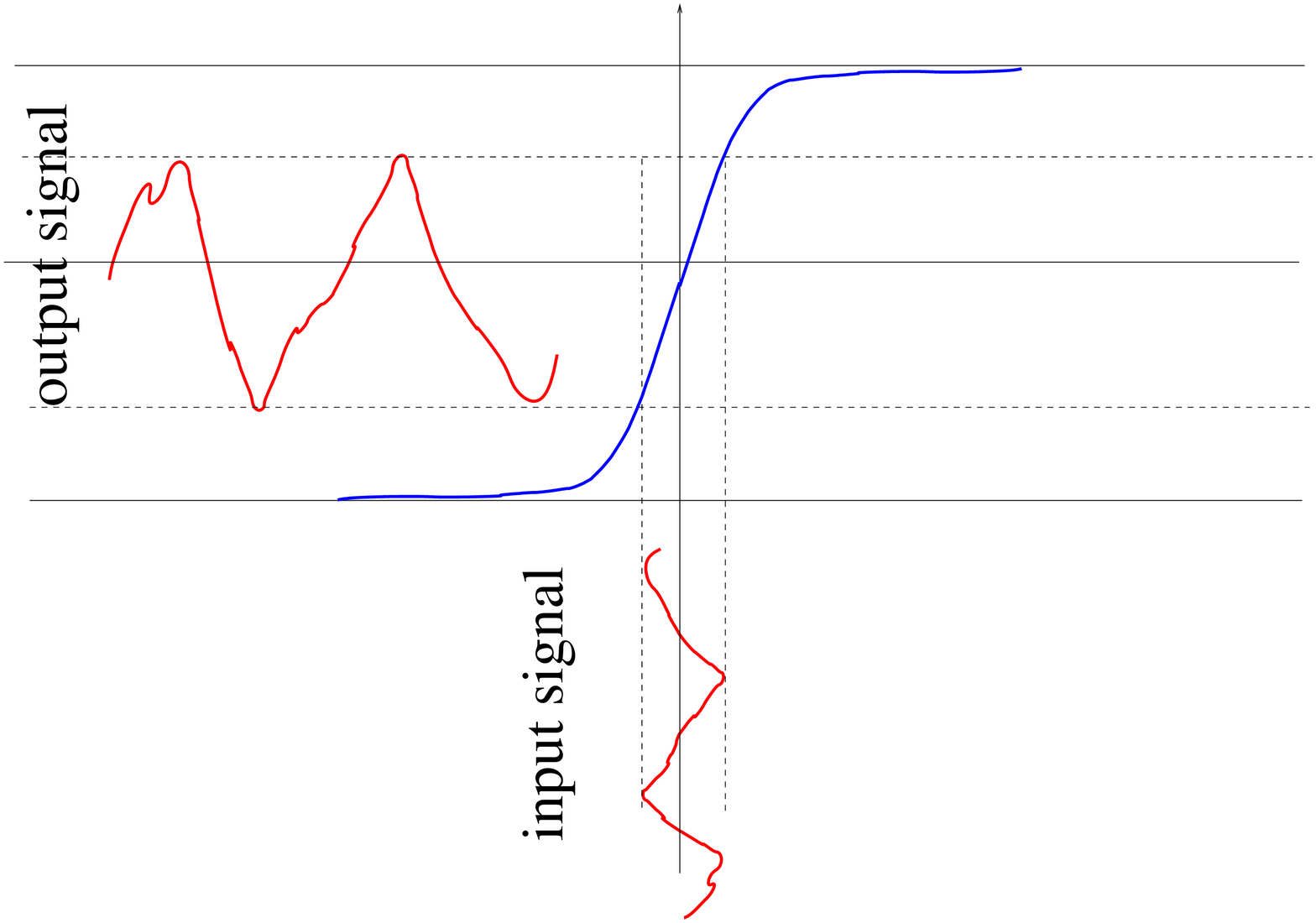}
\hspace{1cm}
\includegraphics[height=5cm,width=5cm,clip=false]{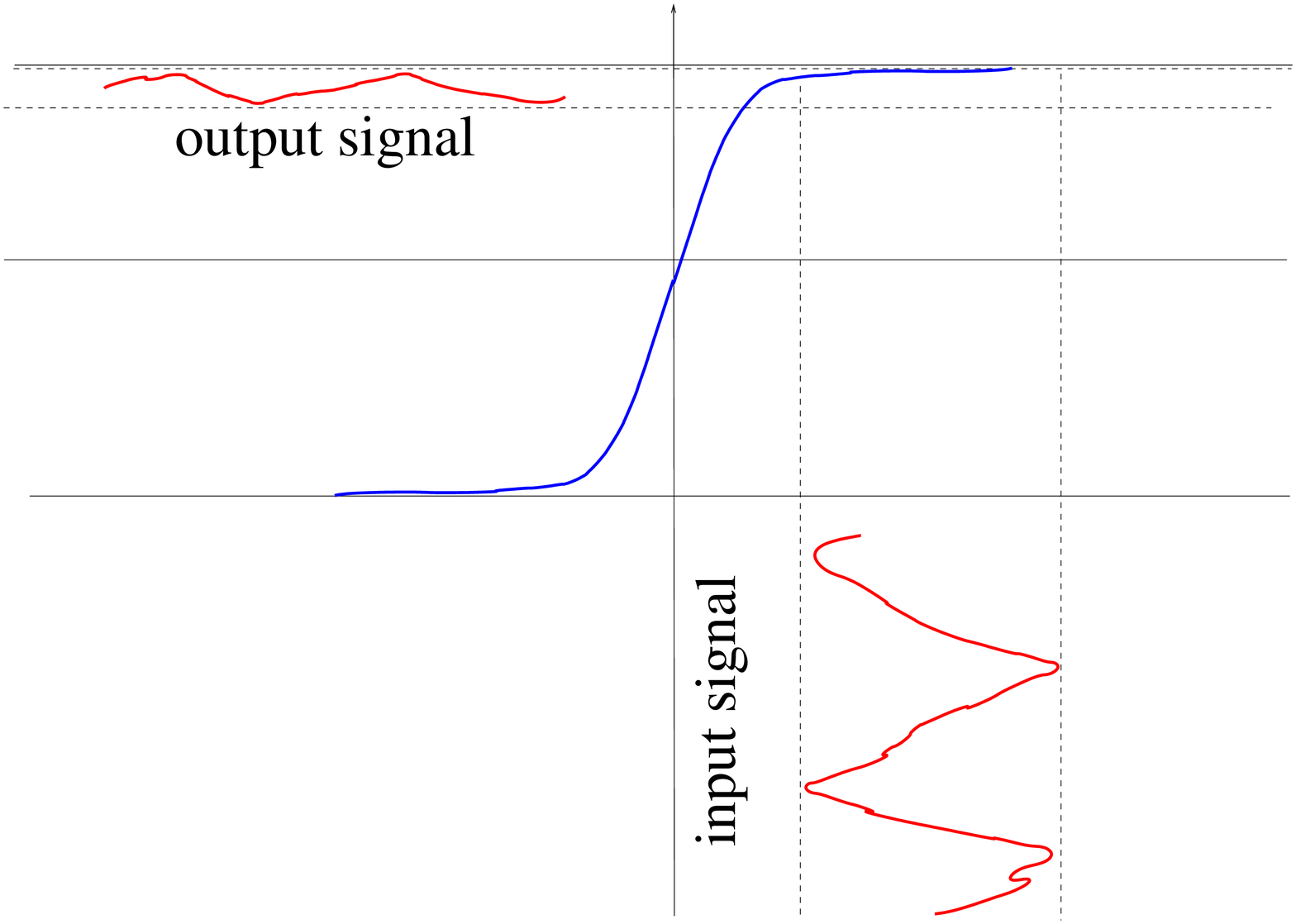}
\caption{\label{FSat} nonlinear effects induced by a transfer function with a sigmoidal shape on signal
transmission. Fig. \ref{FSat}a. Amplification. Fig. \ref{FSat}b. Saturation.}
\end{figure}

Assume indeed that we superimpose to the ``background'' input $u_i(t)$ of the node $i$ a small signal $\xi_i(t)$.
How does this signal propagates inside the network ? Because of the sigmoidal shape of the transfer functions the
answer depends crucially on the activity of the nodes.
 Assume, for the moment and for simplicity, that the time-dependent signal $\xi_i(t)$ has variations
substantially faster than the variations of $u_i$. Consider then the cases depicted in
Fig. \ref{FSat}a,b.
 In the first case the signal $\xi_i(t)$ is amplified by $f$, without distortion if $\xi_i(t)$ is weak enough.
On the other hand, it is damped and distorted by wild nonlinear effects due to saturation, in Fig. \ref{FSat}b. This 
example shows  that the signal propagation in such a network must take into account the topological structure
of the graph \textit{as well as the nonlinear effects}. This simple remark  leads one to reconsider the notion
of ``hub''. A hub is a relay with a strong connectivity. From a topological point of view, this is certainly a very important
node. But, when considering signal propagation in networks with saturating relays, the role of a hub might be
temporarily weakened if this hub is maintained in a saturated state by the global activity.\\
%
%
%
%
%
%
%
%
%
%
%

In some situations, it is possible to analyze the combined effects of topology and nonlinearity on the  propagation of weak amplitude signals.
This is the goal of the method developed in \cite{CS}. On technical grounds one first needs to make the assumption
that the global spontaneous activity is chaotic. The relevance of chaos for realistic situations may be debated (note
however that chaos generically occurs in the model presented below) and we deal with this point in the discussion.
But, at the present stage, our point is slightly different. On one hand
we provide an example where the combined effects  of topology and nonlinearity can be handled, and, on the other hand,
 we establish
that one can study the propagation and the effects of a signal superimposed upon the chaotic background \textit{in spite of}
(and in fact thanks to) chaos.\\

To be more specific consider the following model. The input signal $u_i(t)$ is a function
of the activity $x_j(t)$ of the units $j$ connected to $i$ and it is given by $u_i(t) = \sum_{j} J_{ij} x_j(t)$.
Then the global dynamics writes:
\beq\label{DNu}
\bu(t+1)=\bG\left[\bu(t)\right]=\cJ.f(\bu(t)), 
\eeq
\nid where  $\bu(t) = \left\{u_i(t)\right\}_{i=1}^{N}$ and
where we used the notation $f(\bu(t)) = \left\{f(u_i(t))\right\}_{i=1}^{N}$.
$\cJ$ is the matrix of coupling coefficients.

The saturation of the sigmoid discussed above (fig. \ref{FSat}b)
 has the dynamical effect of producing volume contraction in the phase space. Indeed, the
Jacobian matrix is given by $DG_\bu = \cJ\Lambda(\bu)$, where $\Lambda$ is a diagonal matrix
with  $\Lambda_{ii}(\bu)=f'(u_i)$, and its 
determinant is given by $\det DG_\bu = \det(\cJ) \times \prod_{i=1}^N f'(u_i)$. The determinant has an absolute value strictly
lower than $1$ provided that some $u_i$'s are strong enough (corresponding to a saturation of the corresponding
unit). 

Consequently, the asymptotic dynamics settle onto an attractor. This attractor is generically unique (provided
that one breaks the symmetry $\bu \to -\bu$ of the transfer function $f(x)=tanh(gx)$, with, e.g. a small time independent threshold added to
the local field $\bu$).

The dynamics (\ref{DNu}) generically exhibits a transition to chaos by quasi-periodicity as $g$ increases,
for suitable choices of the $J_{ij}$'s (e.g. independent, identically distributed random variables
with scaled mean and variance  \cite{IJBC,PD,EPL,JP}). [Recall however that the $J_{ij}$'s are fixed during the evolution
and that we consider a specific realization of the $J_{ij}$'s (we do not average over the disorder)].
Thus the dynamics asymptotically settles onto a chaotic attractor provided that $g$ is sufficiently large.
The statistical properties of the dynamics on its attractor are characterized by the Sinai-Ruelle-Bowen measure $\rho$
(SRB) which is obtained as the weak limit of the Lebesgue measure $\mu$ under the dynamical evolution:
\beq \label{SRB}
\rho=\lim_{n \to +\infty} \bG^{n} \mu.
\eeq

In the following we will assume that all Lyapunov exponents are bounded away from zero
(weak hyperbolicity). Then for each $\bu \in supp \rho$, 
where $supp \rho$ is the support of $\rho$, there exists a splitting 
$E^s(\bu) \oplus E^u(\bu)$ such that $E^u(\bu)$, the unstable space, is locally tangent to the attractor (the local unstable manifold)
and $E^s(\bu)$, the stable space, is transverse to the attractor (locally tangent to the local stable manifold).
 Let us emphasize that the stable and
unstable  spaces depend on $\bu$ (while the Lyapunov exponents are $\mu$ almost surely constant).
Let us consider a point $\bu$ on the attractor
and make a small perturbation $\delta_\bu$. This perturbation can be decomposed as
$\delta_\bu = \delta_\bu^u + \delta_\bu^s$ where $\delta_\bu^u \in E^u(\bu)$ 
 and $\delta_\bu^s \in E^s(\bu)$. 
$\delta_\bu^u$ is locally amplified with an exponential rate (given by the largest
positive Lyapunov exponent). On the other hand 
 $\delta_\bu^s $  is damped with an exponential speed (given by the largest
negative Lyapunov exponent).\\
   
Assume now that we superimpose a weak signal upon the (chaotic) activity. For simplicity,
we shall assume that the signal does not depend on the state of the system (linear response
still applies in this case, but the equations (\ref{dro},\ref{Chi}) do not hold anymore). Denote
by $\bxi$ the vector $\left\{\xi_i \right\}_{i=1}^N$. The new dynamical
system is:
\beq \label{pert}
\tbu(t+1)=\bG\left[\tbu(t)\right]+\bxi(t)=\tbG\left[\tbu(t)\right]
\eeq 

 The weak signal $\bxi(t)$ may be viewed as small perturbation of the trajectories of the
unperturbed system (\ref{DNu}). Consequently it has a decomposition on the local stable and unstable space.
The stable component is exponentially damped. The unstable one is amplified by the dynamics and nonlinear
terms  rapidly scramble and mix the signal. Consequently, it becomes soon impossible to distinguish the signal from the chaotic
background.

 This is the effect observed on \textit{individual trajectories}. However, the situation
is substantially different if one considers the \textit{average} effect of the signal, the average being performed with respect
to the SRB measure $\rho$ of the unperturbed system. It has been established in \cite{CS} that
the \textit{average} variation of the local field $u_i$ under the influence
of the signal is given, to the linear order, by:
\beq\label{dro}
\rho\left[\delta_{u_i}(t)\right] \deq \left< \tu_i(t) - u_i(t) \right>=\sum_{\sigma=-\infty}^{t}\chi(\sigma)\bxi(t-\sigma-1),
\eeq
\nid where $\chi(\sigma)$ is the matrix :
\beq\label{Chi}
\chi(\sigma) = \int \rho(d\bu)  D\bG^{\sigma}_{\bu} 
\eeq
\nid that writes in explicit form:
\beq\label{chiij}
\chi_{ij}(\sigma)=
\sum_{\gamma_{ij}(\sigma)}
        \prod_{l=1}^{\sigma}J_{k_l k_{l-1}}
\left< \prod_{l=1}^{\sigma}f'(u_{k_{l-1}}(l-1))\right>.
\eeq
We used the shortened notation $< \ >$ for the average with respect
to $\rho$. The sum holds on each possible path
$\gamma_{ij}(\sigma)$, of length $\sigma$, connecting the
unit $k_0=j$ to the unit $k_\sigma=i$, in $\sigma$ steps.
One remarks that each
path is weighted by the product of a \textit{topological} contribution
depending only on the weight $J_{ij}$ and
of a \textit{dynamical} contribution. Since $f$ is a sigmoid 
the weight of a path $\gamma_{ij}(\sigma)$ depends crucially on
the state of saturation of the units $k_0, \dots, k_{\sigma-1}$ at times $0, \dots, \sigma-1$.
In particular, if $f'(u_{k_{l-1}}(l-1))>1$ a signal is amplified while it is damped if
$f'(u_{k_{l-1}}(l-1)) < 1$ (see Fig. \ref{FSat}). Consequently, though a signal has many choices for going from $j$ to $i$ in $\sigma$ time steps,
some paths may be ``better'' than some others, in the sense
that their contribution to $\chi_{ij}(\sigma)$ is higher. In particular, this contribution depends
strongly on the time correlations between the levels of saturation $f'(u_j),f'(u_{k_1}), \dots, f'(u_i)$ of the units
$j,k_1, \dots, i$ composing the path.\\

 This observation leads us to several remarks.

\ben

\item The paths $j \to i$ are a priori \textit{not equivalent}. They have a different weight that depends on one hand
on the topological contribution and on the other hand on the nonlinearity of the transfer function (this last
effect indeed does not exists for linear transfer functions).

\item The average effect of the signal, measured at time $t$, is a sum of a large number of contributions, resulting from the various possible 
paths, with time delayed versions of the signal. Assuming for example that $\bxi$ is periodic, the observed effect is that 
of a sum of waves with different amplitudes and delays. The global effect can be weak if the waves interfere in a destructive way,
or strong if they interfere in a constructive way. This suggests that \textit{resonances} may occur. If one computes the Fourier
transform of $\chi_{ij}(t)$, denoted by $\hchi_{ij}(\omega)$ and called \textit{complex susceptibility}, one observes resonance
peaks corresponding to poles in the complex plane (see \cite{CS}). An example is given in the section \ref{Ssus}.

\item A natural (physicists) reflex would be to seek these resonance in the Fourier transform of the correlation function
$C_{ij}(t)=\left<u_i(t)u_j(0)\right> -\left<u_i\right>\left<u_j\right>$
of the  pair $ij$. Indeed, the fluctuation-dissipation theorem basically tells us that a susceptibility is
(the Fourier transform of) a correlation function.
This is true for dynamical systems encountered in physics, where the (microscopic) dynamics preserves the volume in the phase
space (Liouville theorem).
 But this is no longer true in our case where the dynamics contracts the phase space volume. Actually, the complex susceptibility
\textit{contains more information than the power spectrum}.  

Indeed, the Jacobian matrix, as we saw, can be split into 2 parts, corresponding to the action of $\bG$ in
the local stable and unstable space,  respectively. This means that the response function (\ref{susth}) (the corresponding susceptibility)
decomposes in a stable and an unstable part. Each part has its resonances and they can be drastically different. On the one hand
it has been shown by Ruelle \cite{Ruelle} that the unstable contribution is actually a correlation function (this is a generalized version of the
fluctuation-dissipation theorem). Henceforth the resonances of the unstable part are contained in the power spectrum.
They are called ``Ruelle-Pollicott resonances''
\cite{RP} and \textit{they do not depend on the observable} (provided the observables belong to the same
suitable functional space). Practically, in our case, this means that these resonances do not depend on the pair $ij$
\footnote{More precisely the pole, whose real part is the frequency value of the maximum, and whose imaginary part is
the resonance width, does not depend on the pair, but the residue, corresponding of the value of maximum, depends on it (see e.g. Fig.
\ref{FCorrelations})}.
Furthermore, since $|\hC_{ij}(\omega)|=|\hC_{ji}(\omega)|$ the analysis of these resonances
does not tell us which units excites and which unit responds (see section \ref{Ssusvscor}). In this sense
the analysis of the correlation does not display \textit{causal} information (except the trivial
property $C_{ij}(t)=C_{ji}(-t)$).
But the main drawback of correlations functions is that \textit{they do not contain all possible resonances}.

Indeed, the \textit{stable part} displays additional resonances, called in the following \textit{stable resonances}. They are
 not Ruelle-Pollicott  resonances. They  may also depend on the pair $ij$.
Indeed, the susceptibility of the pair $i \to j$ is in general distinct from the susceptibility of the pair
$j \to i$ and they may have distinct resonances (see Fig. \ref{FReponse}a).
Note also that the corresponding response function are \textit{causal}
 since the stable directions introduce an arrow of time (see Fig. \ref{FReponse}b).
Finally, on numerical grounds, the computation of complex susceptibilities affords a better resolution frequency than the computation
of correlation functions (see section \ref{Ssusvscor}).

 Therefore the complex susceptibilities give us essential information about the average  
effect of a periodic signal, applied by some unit onto some other unit.
This is investigated in some details in sections \ref{Ssus},\ref{Smodamp}.

\item The equation (\ref{chiij}) opens in principle the possibility of inducing a response of $i$, by exciting $j$ with a suitable
frequency, \textit{even if there is no direct link between the 2 units}. On the contrary, there may exist a direct link between
$j$ and $i$ and, in spite of this, there may not be a measurable effect if the frequency of the excitation applied to $j$ does not correspond to a high
response of $i$. This enhances the effect of nonlinearities in the effective capacities of the network. This is discussed in the section
\ref{Ssus}. The notion of ``hub'' is in particular revisited in the section \ref{SHub}.

\item The existence of the stable part may lead to a violation of the standard wisdom stating that the characteristic time of return
to equilibrium is equal to the characteristic time for \textit{mixing}. Actually, stable resonances introduce additional time scales
that can be relatively longer than the mixing time. (The mixing time is given by the Ruelle-Pollicott resonance that is the closest to the real axis in the
complex plane). As a matter of fact, it is in principle possible to observe the average effect of a signal corresponding to a kick, over
a time substantially larger than the correlation time. An example is shown in section \ref{Ssusvscor}.

\item Finally, the existence of resonances opens the possibility of using \textit{amplitude modulation} to transmit a signal from a specific
unit to another specific one, in spite of (thanks to) chaos. The original signal is then recovered by a suitable averaging procedure.
This is discussed in section  \ref{Smodamp}. Let us emphasize that the procedure suggested here \textit{is not control of chaos}.
We do not stabilize the dynamics on a periodic orbit by a  suitable perturbation. The perturbed dynamics stays chaotic and we
use some natural properties of chaos, such as mixing, to reconstruct \textit{any} weak signal by a suitable average.
 
\een

In the next section, we present an example, based on the model (\ref{DNu}), supporting these claims. 
For this we select a specific set of $J_{ij}$'s, randomly drawn, and we focus on the characteristics
of this \textit{particular} network. \textit{We do not perform statistical averages on the distribution of the $J_{ij}$'s}. 
 As argued in the introduction we want indeed to provide analysis tools
allowing an user to extract the characteristics of the network \textit{he is currently using}.  
One may however ask
about the genericity of this example.   What happens for a different set of $J_{ij}$'s ? What happens if one
increases the size ? The statistical behavior of this model has been widely studied in \cite{IJBC,PD,JP}. It has
 been shown that chaos generically occurs   provided $g$ is sufficiently large.  The average critical value for the 
transition to chaos has been analytically computed in \cite{EPL,JP}.  The thermodynamic limit was also fully characterized
for a mean field version. From these studies and from the theoretical arguments developed above we claim that
the behavior described is generic in this model. Another example of resonance curves has been produced
in \cite{CS} for a fully connected version of the coupling matrix. 

Obviously, some features such as the resonance frequencies
are \textit{specific} to the choice of the $J_{ij}$'s. But this is precisely what interests us.
Starting from a chaotic network with specific resonances, we are able to compute numerically
the susceptibilities by suitably exciting the nodes. This procedure does not require an a priori
knowledge about the dynamics. From the susceptibilities curve we extract the resonances of this network
and we then use them for applications.

One may also ask about the genericity of the \textit{model} itself. This point is examined in the discussion. 
At this stage we simply want to remark that most of the effects exhibited in the next section
are predicted from the general theory presented above. These effects are non-intuitive and depart
widely from the conventional wisdom about chaotic systems. They also open new perspectives in the
study of networks. This example is therefore designed to check that these theoretical predictions
can be realized in at least \textit{one} example. Also, the analysis performed here can be easily
reproduced in other examples or models, possibly  more realistic (see the discussion).
 
\section{Signal propagation} \label{SignProp}

\ssu{Model example} \label{SModel}

The numerical simulations presented here have been performed on the following example. 
The number of units was fixed to $N=9$.  The network is sparse. Each unit receives connection
from exactly $K=4$ other units.
The $J_{ij}$'s have been drawn at random according to a Gaussian distribution with
mean zero and a variance $\frac{J^2}{K}$. This ensures the correct normalization of the
local fields $u_i$ \cite{IJBC}. The version of the $\cJ$ for which all simulations have
been performed is : 

\bigskip

$$
\tiny{
\left[
\baR{ccccccccccccccccc}
0 & 0 & 0.213 & 0 & 0.469 & 0 & 0 & 0.69 & 0.318 \\
 -1.131 & 0.822 & 0 & 0 & 0 & 0 & 0 & 0.007 & -0.301 \\
 0 & -0.234 & 0 & 0 & -0.51 & -0.283 & -0.177 & 0 & 0 \\
 0 & 0.644 & 0 & 0 & 0.033 & 0 & 1.187 & 0.722 & 0 \\
 0 & 0 & 0 & 0 & 0.511 & -0.579 & -0.495 & 0.269 & 0 \\
 0 & 1.015 & 0 & 0 & 0 & 0 & -1.312 & 0.684 & -0.365 \\
 0 & 0 & -0.852 & -0.342 & 0.389 & 0 & 0 & -0.041 & 0 \\
 0 & 0.416 & 0 & 0 & -0.084 & 0 & 0.287 & 0.208 & 0 \\
 0 & 0 & 0 & -0.649 & 0 & 0 & 0.331 & 0.140 & -1.023
\eaR
\right]
}
$$

\vspace{0.5cm}

\bc
\footnotesize{Coupling matrix $\cJ$ used in the simulations below.}
\ec

\bigskip

\begin{figure}[!ht]
\includegraphics[height=6cm,width=6cm,clip=false]{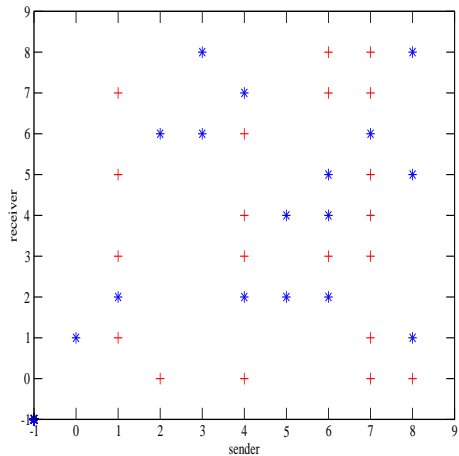}
\caption{\label{FJij} Connectivity matrix for the simulations performed in this paper.}
\end{figure}

(Note that the corresponding graph is not decomposable).
The corresponding network is drawn in Fig. \ref{FJij}. Blue stars correspond to inhibitory links
and red crosses to excitatory links. It is for example easy to see that the unit $7$ is a ``hub'' in
the sense that it sends links to almost every units, while $0$, $2$, $3$ or $5$ send at most two links.

%
%
%

A small constant $\theta_i$ has been added to each $u_i$ to break down the symmetry $\bu \to -\bu$ (i.e. $u_i(t)=\sum_j J_{ij}x_j(t)+\theta_i$).

The corresponding dynamics exhibits a transition to chaos by quasi-periodicity. For $g=3$
the dynamics has a strange attractor. There is one positive Lyapunov exponent ($\lambda_1=0.153$)
and $8$ negative Lyapunov exponents (with $\lambda_2=-0.427$). Hence the system is weakly
hyperbolic (all Lyapunov exponent bounded away from zero). The spectrum is stable to small variations of $g$.
The Kaplan-Yorke dimension is  $1.64$. 

\ssu{Computation of susceptibilities.}\label{Ssus}

In a nutshell (see \cite{CS} for more details)
the computation of susceptibilities  consists in perturbing the trajectories of (\ref{DNu}) by two perturbations
  $\bxi^{(1)}(t)= \epsilon \be_j \cos(\omega t)$
and  $\bxi^{(2)}(t)= - \epsilon \be_j \sin(\omega t)$. If 
$\tbu^{(1)},\tbu^{(2)}$ denote the variables of the corresponding perturbed systems
then one may write (for $\omega \neq 0$):
\beq\label{susth}
\hat{\chi}_{ij}(\omega) =\lim_{T\to\infty}{1\over T\epsilon}\sum_{t=0}^T e^{i\omega(t-1)}\,
[  \tu_i^{(1)}(t)  + i  \tu_i^{(2)}(t )]. 
\eeq
This provides 
a straightforward way to compute
the susceptibility where most of the computing time 
 goes into computing the orbits $\tbu^{(k)}(t)$. From a numerical point of view the precision of (\ref{susth}) can be improved by
performing an additional average over several trajectories. This allows one to compute error bars.\\

 Note that the average is directly performed on the perturbed trajectories,
and not on the difference between the perturbed and unperturbed trajectories. This has two consequences. On one hand
this avoids to iterate simultaneously the perturbed and unperturbed system, compute the difference, and renormalize it 
when it becomes too large, to keep only linear effects. Instead, the computation (\ref{susth}) includes
all nonlinear effects and these are \textit{precisely} these effects that permit to compute the average
with respect to $\rho$.
Indeed, the SRB measure (\ref{SRB}) is exactly an average over a typical trajectory including all nonlinear effects 
such as mixing and folding.

 The perturbation  $\bxi^{(i)}(t)$ decomposes into a stable and unstable part.
The unstable part is rapidly amplified by the initial condition sensitivity, then nonlinear contributions arise, leading
to mixing and to an effective average \textit{on} the attractor, when $T \to \infty$.
 This average is the Fourier transform of a correlation function
at the frequency $\omega$. Since correlation functions decay exponentially in chaotic system \footnote{The exponential
decay can be proved if the system is uniformly hyperbolic but uniform hyperbolicity can not be checked numerically.
Henceforth, one assumes that the system behaves as if it ``were'' uniformly hyperbolic. Note however that uniform hyperbolicity
is a sufficient but not a necessary condition for exponential decay. Without exponential decay the sum  (\ref{susth})
may diverge leaving us without linear response theory. Consequently,
on practical grounds, one has to check that the sum (\ref{susth})
does not diverge.} the sum (\ref{susth}) converges and gives a \textit{finite} contribution to the unstable part.
The stable part is damped by contraction, but since it is applied in a continuous way,
one obtains an effective summation of the effects of a sinusoidal perturbation \textit{transverse} to the attractor.
One finally obtains a finite quantity giving the (average) response of the system to a perturbation having projections
in the stable and unstable directions. When
$\epsilon$ is small this is the \textit{linear} response.
Note however that there is a priori no condition on $\epsilon$ in eq. (\ref{susth}) \footnote{There is in fact an hidden one: $\epsilon$ must not be too large to
ensure that the perturbed system is still chaotic. Indeed, clearly a too big $\epsilon$ will irremediably kill the chaos
and give rise to a periodic regime.}.
  This  means that $\hat{\chi}_{ij}(\omega)$ corresponds to the \textit{response} of the system to the
perturbation, possibly including nonlinear contributions in $\epsilon$, whenever $\epsilon$ is too large. One has therefore to check
that $\epsilon$ is small enough to ensure that $\hat{\chi}_{ij}$ does not vary when $\epsilon$ varies on a small interval.
 An example is given in section \ref{Ssusvseps}. \\

\begin{figure}[ht!]
\includegraphics[height=5cm,width=14cm,clip=false]{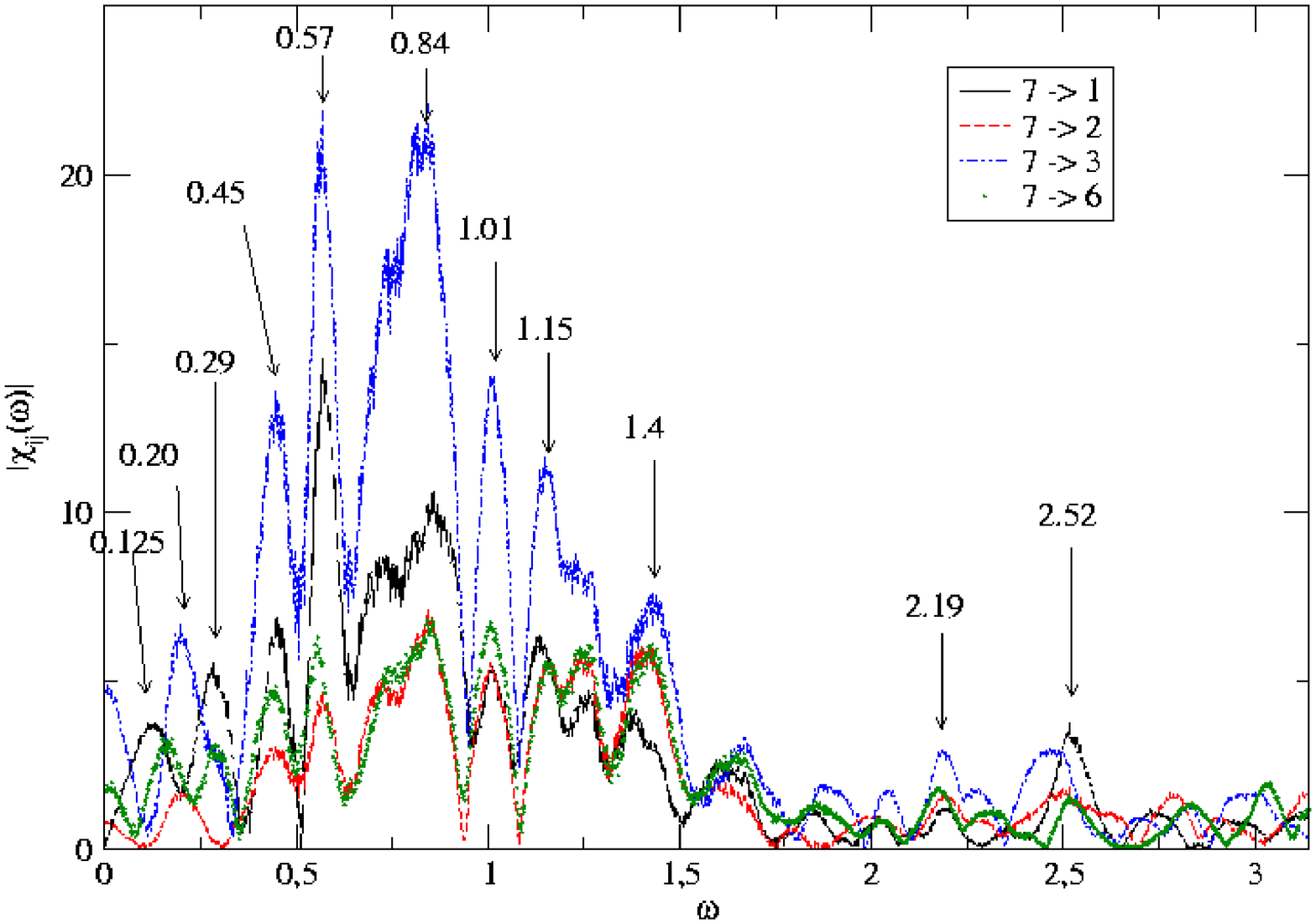}
\vspace{0.2cm}
\includegraphics[height=5cm,width=14cm,clip=false]{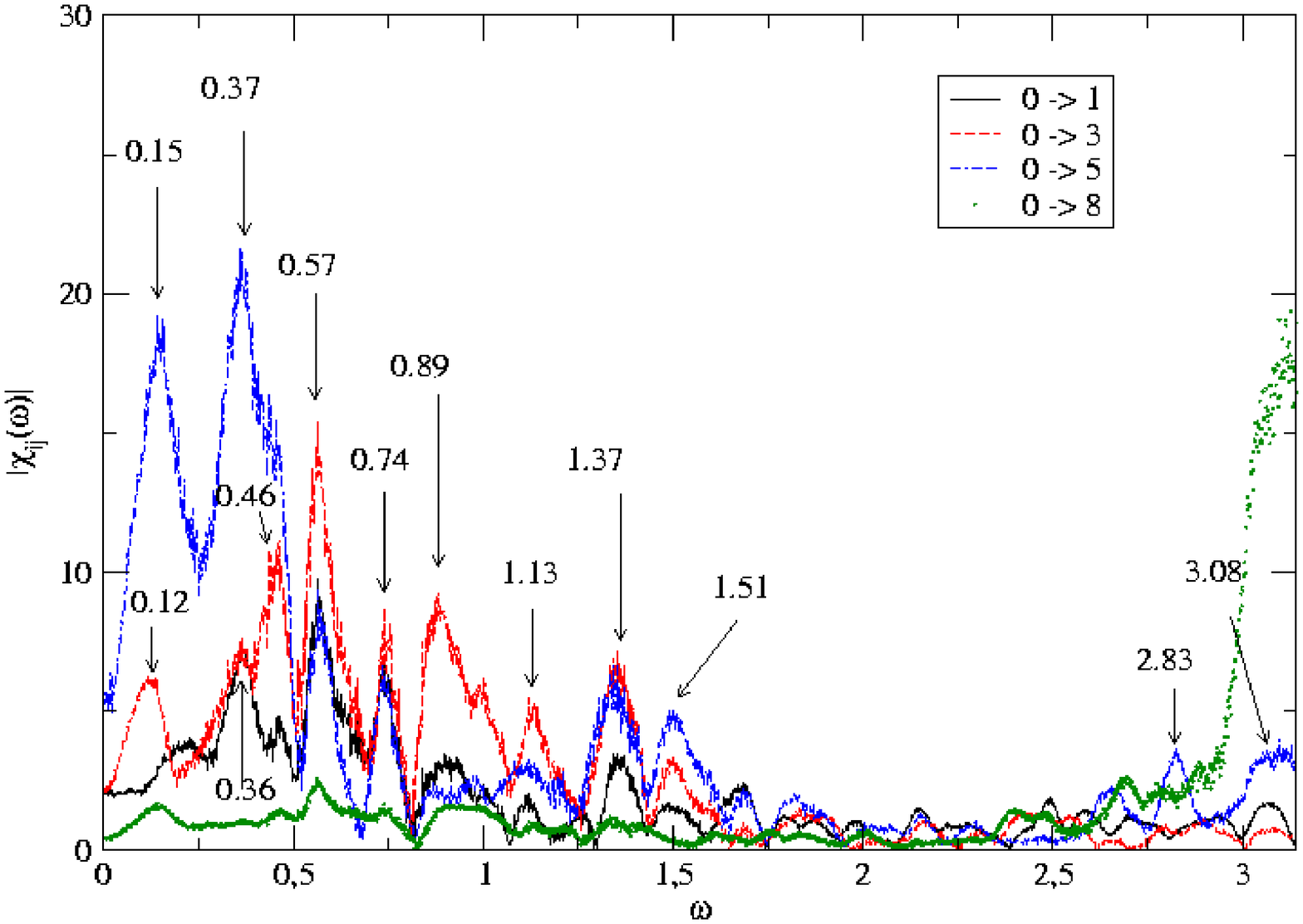}
\vspace{0.2cm}
\includegraphics[height=5cm,width=14cm,clip=false]{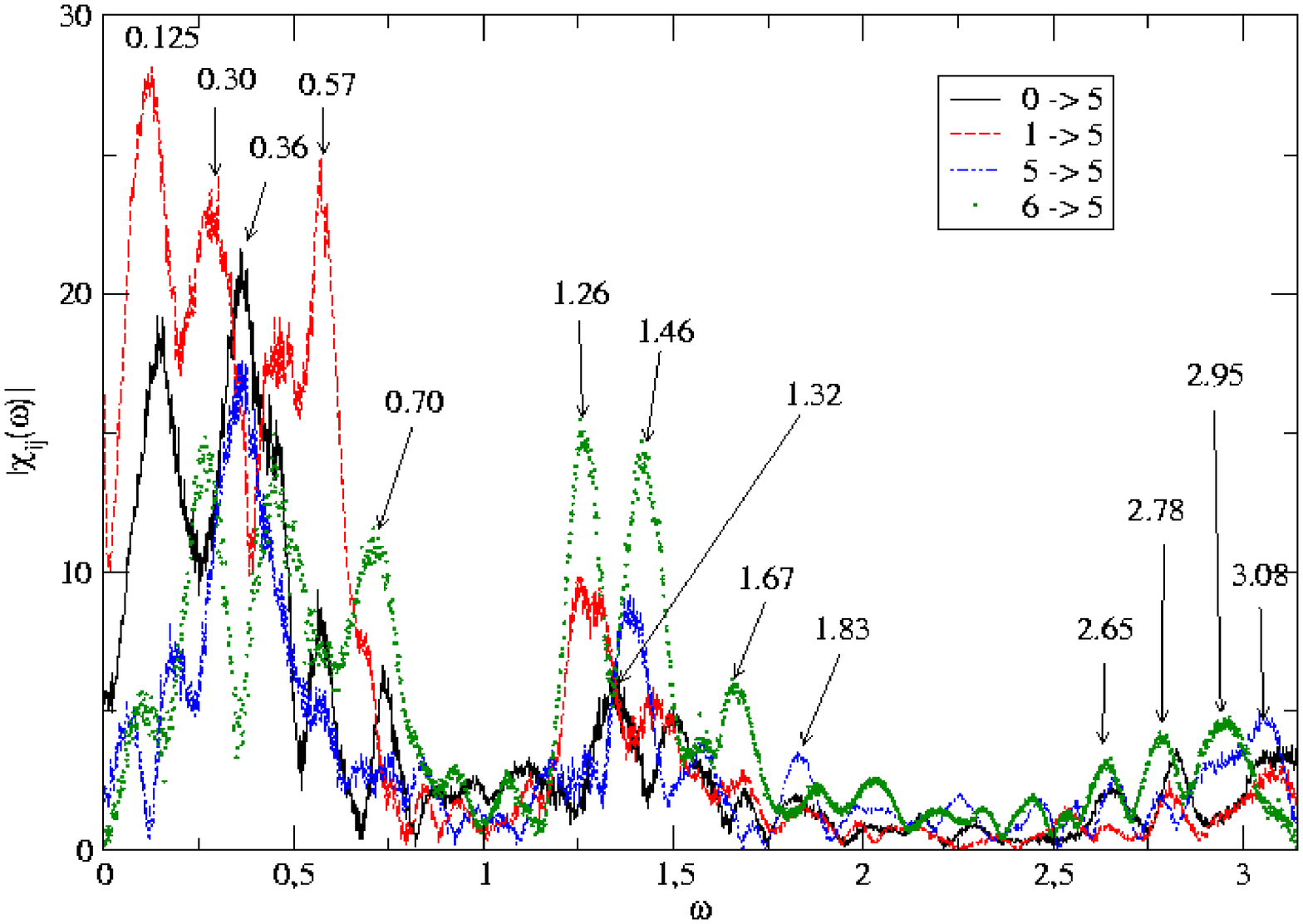}
\caption{\label{FResonances}. Modulus of some susceptibilities. Fig. \ref{FResonances}a. 7 (highly connected unit) excites the units:
1 (excitatory link with intensity $J_{17}=0.007$); 2 (no direct link); 3 (excitatory link with intensity $J_{37}=0.722$);
6 (inhibitory link with intensity $J_{67}=-0.041$). Fig. \ref{FResonances}b. 0 (weakly connected unit) excites the units:
1 (inhibitory link with intensity $J_{10}=-1.131$); 3,5,8 (no direct link); Fig. \ref{FResonances}c. 5 receives the excitation
from the units:  0 (no direct link); 1 (excitatory link with intensity $J_{51}=1.015$); 5   (no direct link); 6 
(inhibitory link with intensity $J_{56}=-1.312$).}
\end{figure}

Some examples of susceptibilities computed in this way are depicted in Fig. \ref{FResonances}
with $\epsilon=10^{-3}$. The computation has been done with $T=262144$ and $100$ trajectories
corresponding to have $26.214.400$ points for each $\omega$. The frequency sampling is $\delta \omega =
\frac{2 \pi}{4096}$.
%
%
%

\begin{figure}[!ht!]
\includegraphics[height=8cm,width=12cm,clip=true]{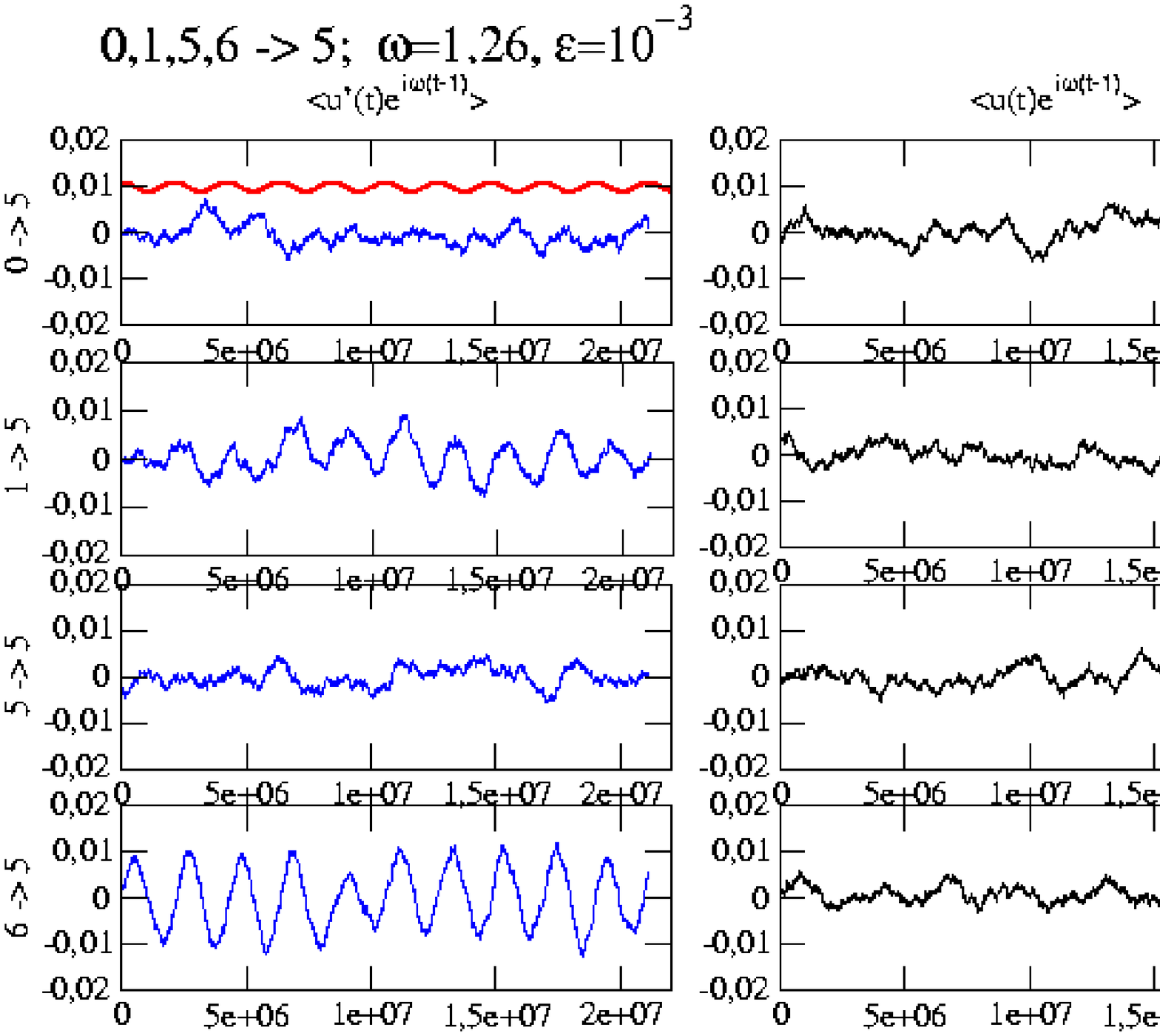}
\caption{\label{FReponse_5_Exc_all_om1.26}  Response of unit $5$ to an excitation of the units $0,1,5,6$ with
amplitude modulation. The fundamental frequency is $\omega_0=1.26$ and the modulation frequency is $\omega_M=2.97 10^{-6}$.}
\end{figure}

Several remarks can be made. First, as expected, there are resonance peaks common to all pairs. For example,
 there is a common  peak located at  $\omega=0.57$ with the same width (corresponding to the imaginary part of the corresponding pole)
Note however that the height can be different (it corresponds to the value of the residue).  
 It is also clear from inspection of Figure \ref{FResonances} that there are resonance peaks 
common to the susceptibilities corresponding to the same \textit{emitting} unit (e.g. $0.45;0.84;1.01$ in Fig.\ref{FResonances}a;
$0.74;1.37$ if Fig. \ref{FResonances}b).
 There are also
peaks that exists only for some pairs (e.g. $0.125;0.2;2.52$ in Fig.  \ref{FResonances}a; $0.12,0.74,2.83$ in  Fig.  \ref{FResonances}b;
$0.125,0.70;1.26$ in  Fig.  \ref{FResonances} c).
 Also, a simple glance at Fig.
 \ref{FResonances}c shows  that characteristics of some of the resonance peaks 
(namely the frequency corresponding to the maximal response \textit{and/or} the width)
 of a receiving  unit  depend on the \textit{exciting unit}. For
example, applying a signal to the unit $6$ with the frequency $\omega=1.26$ will induce a strong response of the unit
$5$, while applying the same signal with the same frequency to the units $0$ or $5$ will induce a weak response (see section \ref{Ssus}
and Fig. \ref{FReponse_5_Exc_all_om1.26} for more details). 
  This suggests  that a suitable filtering of the global signal arriving at $5$ will produce a good signal to noise ratio
in the case $6 \to 5$ while it will be poor in the case $0 \to 5$ or $5 \to 5$ (in this last case the unit does not ``feel''
the signal even if it is applied to itself, because this signal is hidden into the chaotic background). 
Note finally that some resonance peaks are relatively high ($\sim 20$) corresponding to an efficient
amplification of a signal with suitable frequency.

It is also clear from these figures that the intensity of the resonance has no direct connection with the intensity or the sign
of the coupling and is mainly due to nonlinear effects. For example, there is no direct connection from $0$ to $3$ or $5$
but nevertheless these units react  strongly to a suitable signal injected at  unit $0$.  On the other hand, there certainly exists a link
between  the resonance curves and the topological connectivity of the node: $7$ is a topological ``hub'' that sends links to almost every units,
and the response curves of the units are rather similar. On the contrary, $0$ sends a unique link to $1$. Consequently, the resonance curves
for the other units correspond to ``indirect'' paths and they look different.        
 
\ssu{Susceptibilities versus correlations.}\label{Ssusvscor}

\begin{figure}[ht!]
\includegraphics[height=5cm,width=14cm,clip=false]{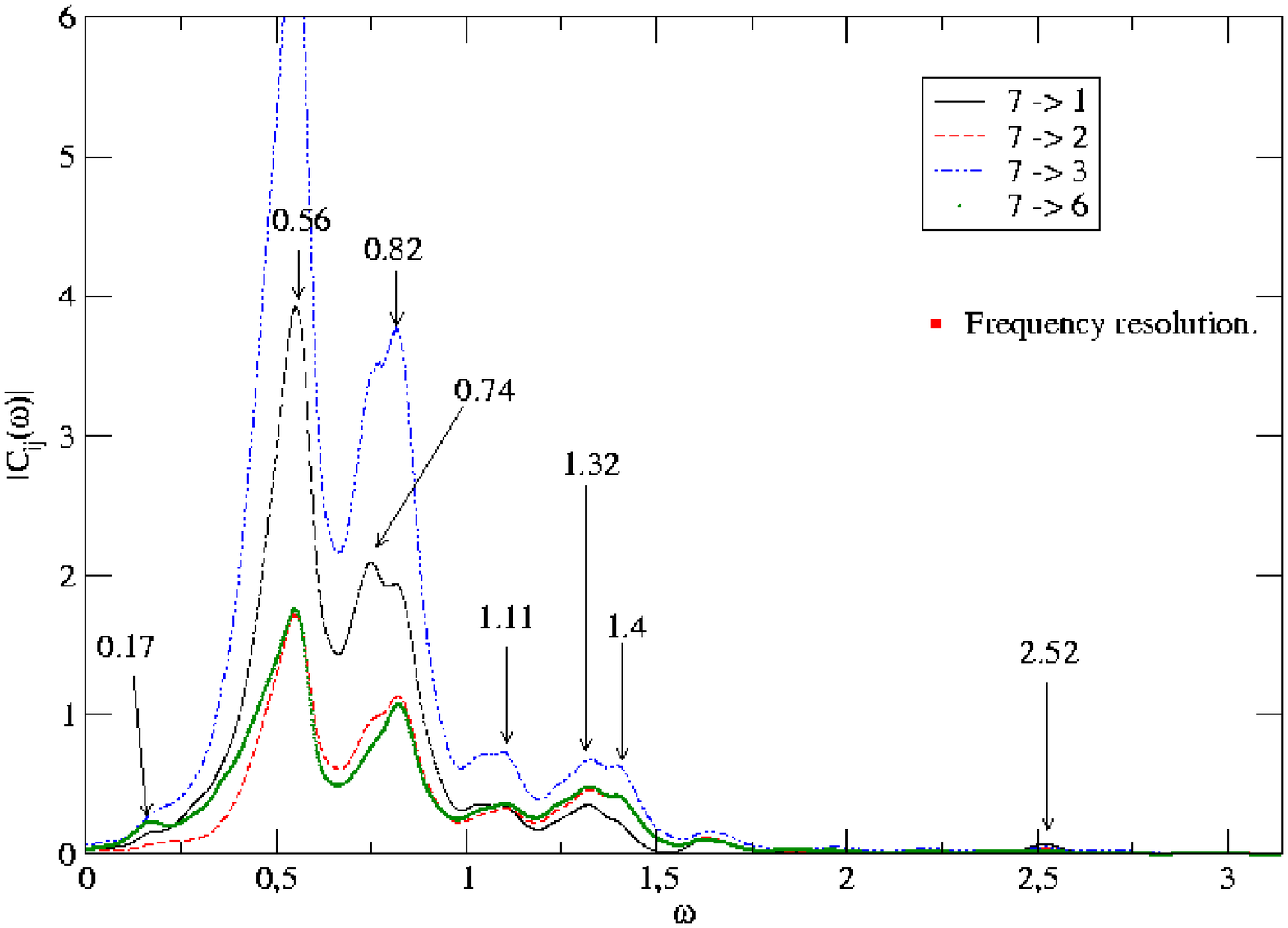}
\vspace{0.5cm}
\includegraphics[height=5cm,width=14cm,clip=false]{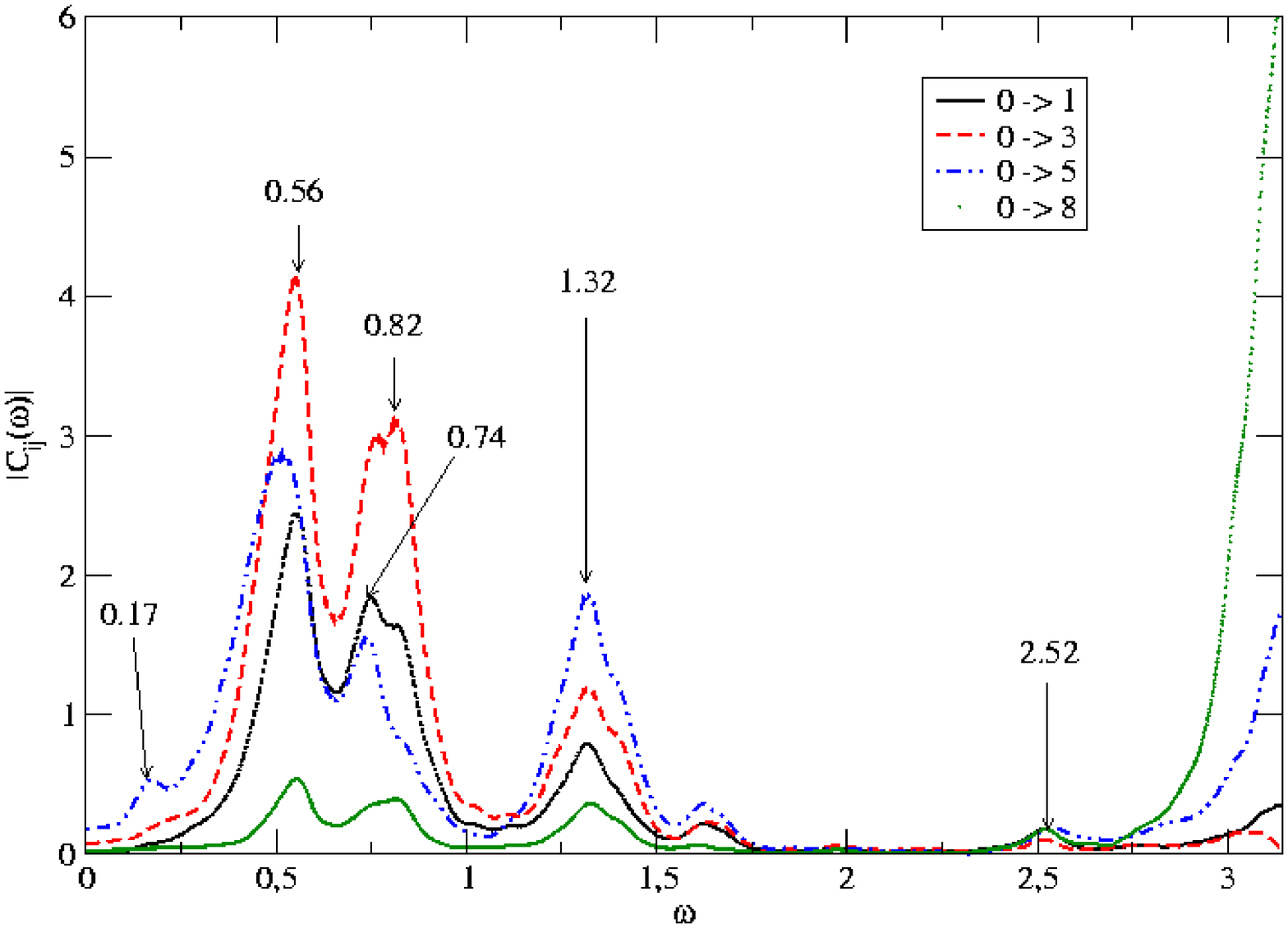}
\vspace{0.5cm}
\includegraphics[height=5cm,width=14cm,clip=false]{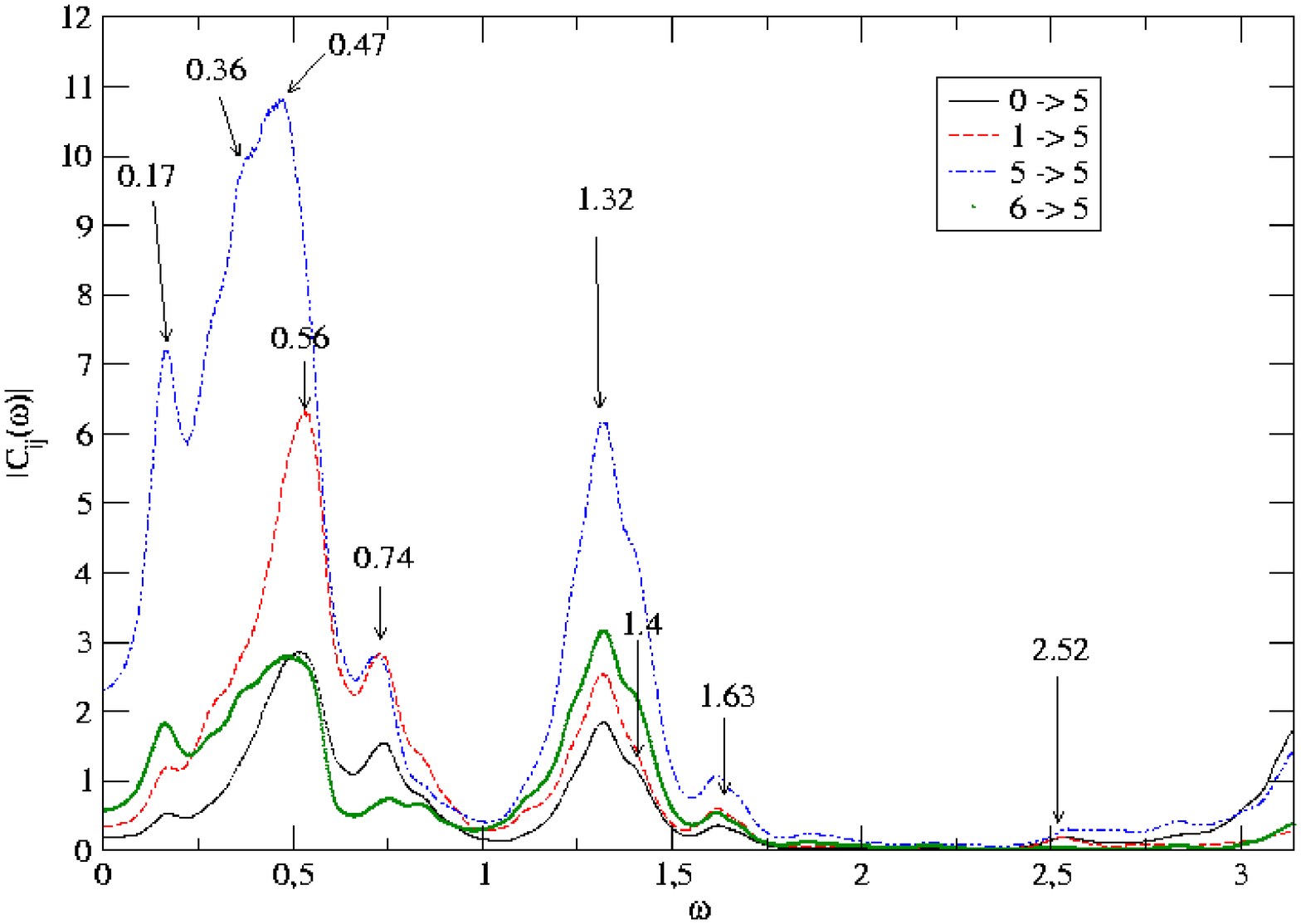}
\caption{\label{FCorrelations}. Modulus of the correlation functions corresponding to the susceptibilities
represented in the Fig. \ref{FResonances}a,b,c. }
\end{figure}

As discussed above, Ruelle's theory states that the susceptibility $\hchi_{ij}(\omega)=\hchi_{ij}^s(\omega)+\hchi_{ij}^u(\omega)$
where $\hchi_{ij}^s(\omega)$ ($\hchi_{ij}^u(\omega)$) is the stable (unstable) part. Consequently,
$\hchi_{ij}(\omega)$ contains stable and unstable resonances. Since unstable resonances are present in the Fourier
transform of the correlation function $\hC_{ij}(\omega)$ it is natural to compare $\hchi_{ij}(\omega)$ and $\hC_{ij}(\omega)$. 
An example is given in Fig. \ref{FCorrelations}. As expected, one observes common peaks  but there are additional peaks in 
the susceptibility. Note also that the correlation curves are all similar and have the same peaks (only the value of the maxima change).

%
%
%

This figure calls however for an important remark. While the numerical method used for the 
computation of the susceptibilities allows one to have a rather high
resolution in frequency  ($\delta\omega_{min} = \frac{2 \pi}{4096}$) and to detect narrow resonance peaks, the computation of the
correlation function is submitted to much more stringent limitations. Indeed, it is well known
that, due to the initial condition sensitivity, the maximum time resolution is (assuming an attractor
with a diameter of order $1$) $t_{max} = -\frac{1}{\lambda_1} \ln(\eta)$, where $\lambda_1$ 
is the maximum Lyapunov exponent ($\lambda_1=0.153$ in our case) and $\eta$ is the round off error
($\sim 10^{-16}$ on a Pentium, in double precision) \cite{ER}. This gives  a $t_{max}$ of order $240$ corresponding
to a frequency resolution $\delta\omega_{min} \sim \frac{2 \pi}{240}=0.026$. This is the narrowest width
of the resonance peaks that one can measure. Using specific libraries in quadruple precision ($\eta \sim 10^{-32}$)
will only divide by 2 the  frequency resolution.
Since this effect is due to initial condition sensitivity in the unstable directions, 
the stable part of the susceptibility is not subjected to these limitations. A consequence of this remark is however that
by  glancing at the example presented here one cannot say whether
  a resonance peak corresponds to a stable or an unstable resonance (except for some striking cases such as $\omega=0.57$).
This would require a more careful investigation of the corresponding poles, but this is not necessary for the scope
of the present work 
(see \cite{CS} for a computation of the poles). Indeed,
 a simple glance at Fig.   \ref{FResonances}a,b,c reveals a large number of peaks and many of them are not
present in the correlation curves. For the applications discussed in the following,
 all what we need to know is that the susceptibility  contains \textit{all}
resonances (stable and unstable) while the correlation  only contains unstable resonances. \\ 

Another observation further underlines the difference
between susceptibilities and correlation functions. The fluctuation-dissipation theorem of non-equilibrium statistical
physics asserts that the response function to a non-equilibrium perturbation can be expressed in terms of
a correlation function. Consequently, the information about non-equilibrium relaxation is included in the equilibrium
fluctuations. It follows in particular that the relaxation time towards equilibrium is equal to the decorrelation time
(mixing time).

\begin{figure}[ht!]
\includegraphics[height=10cm,width=15cm,clip=false]{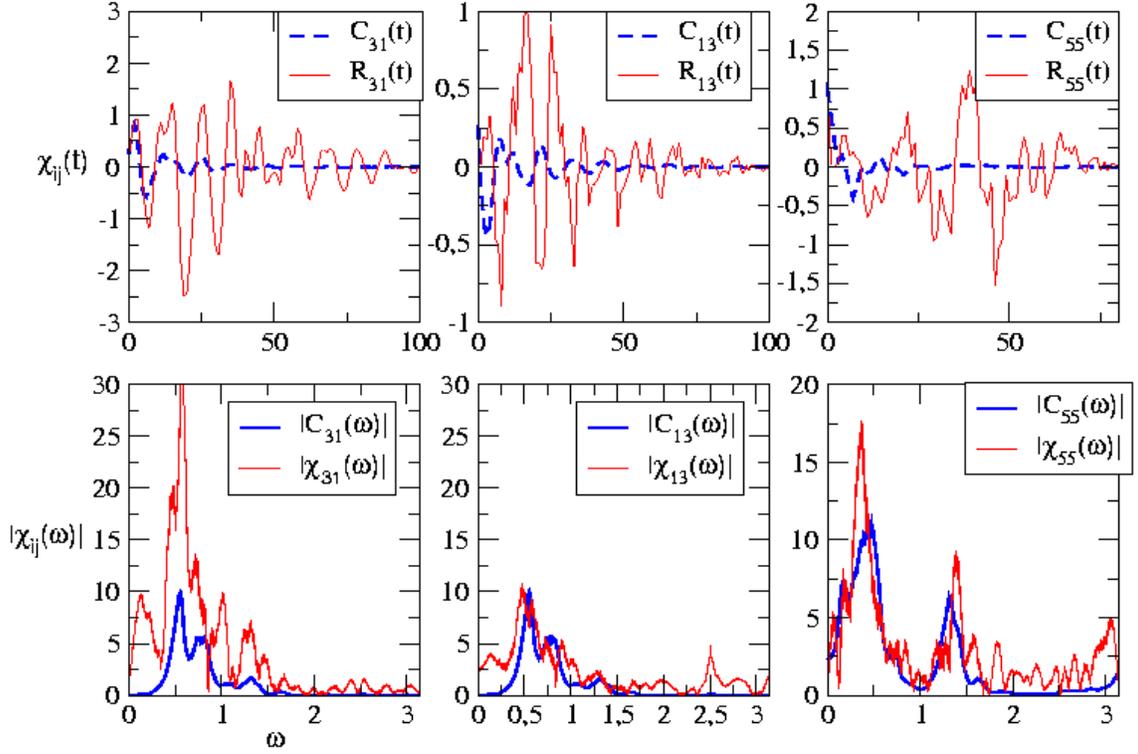}
\caption{\label{FReponse} First row. Average response to a kick compared to correlation function.
Second raw. Comparison of susceptibilities/correlation functions for the pairs $1 \to 3$,
$3 \to 1$, $5 \to 5$.}
\end{figure}

 Consider now  figure   \ref{FReponse}.
The Fourier transform of the susceptibility $\hchi_{ij}(\omega)$ is the average response $R_{ij}(t)$
of $i$ to an instantaneous kick  applied to $j$ at $t=0$. The first row of Fig.  \ref{FReponse} shows these
responses for the pairs $1 \to 3$, $3 \to 1$ and $5 \to 5$ as well as the corresponding time correlations.
Clearly, the \textit{coherence time observed in the response is substantially longer than the correlation time}.
Henceforth, the time for returning to equilibrium is \textit{different}, and in our case longer, than
the mixing time.  This  shows that the (average) effect of a kick can
be observed on very long time scales \textit{in spite of chaos}. 

One also notes that the response $1 \to 3$ is drastically different from the response  $3 \to 1$ while
correlation functions are identical (up to the symmetry $C_{13}(t)=C_{31}(-t)$). This difference is even
more striking when observing the Fourier transforms.
Since $C_{ij}(t)=C_{ji}(-t)$
 the graph of $|\hC_{13}(\omega)|$ and   $|\hC_{31}(\omega)|$ are  identical. Consequently, observing a resonance
peak in the Fourier transform of the correlation function for a pair $ij$ does not tell us ``who excites whom''.
  On the other hand,   the graph of
$|\hchi_{13}(\omega)|$ and $|\hchi_{31}(\omega)|$ displays clearly different peaks. At a frequency $\omega=2.52$, $3$
excites $1$, but $1$ does not excite $3$. Consequently, the susceptibility provides
\textit{causal} informations contrary to correlation functions. The difference comes from the fact
that correlation functions deal with the dynamics ``on'' the attractor, while susceptibilities consider
perturbations on the attractor as well as transverse to the attractor. As a matter of
fact, the presence of stable directions introduces an explicit arrow of time and causality.

%
%
%

\ssu{Revisiting the notion of ``hubs''.} \label{SHub}

\begin{figure}[ht!]
\includegraphics[height=8cm,width=10cm,clip=false]{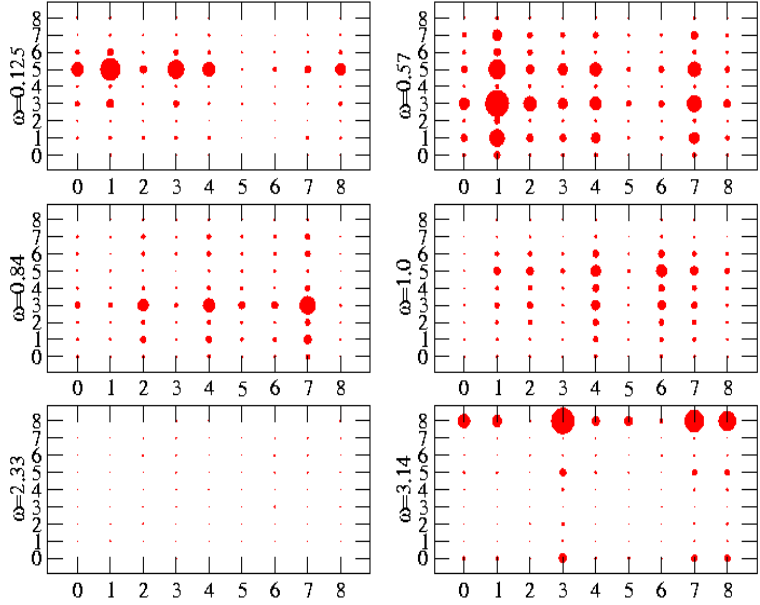}
\caption{\label{FConn} Effective connectivity for $\omega=0.125; 0.57; 0.84; 1.0; 1.26; 2.33 $.}
\end{figure}

The resonance curves leads us to seriously revisiting the notion of hub. As indicated in Fig. \ref{FJij}, the node
$7$ is a topological hub. However, its ability to propagate a weak periodic signal with frequency $\omega$ \textit{depends
on} $\omega$. The previous analysis leads then us to propose a notion of ``effective'' connectivity
based on susceptibility curves. For a  given  frequency $\omega$, we plot the modulus of the susceptibility
$|\chi_{ij(\omega)}|$ with a representation assigning to each pair $i,j$ a circle whose size is proportional
to the modulus. Some examples are represented in Fig. \ref{FConn}. We clearly  see in this figure that changing the
 frequency changes the effective
network.

For example, with a frequency $\omega=0.125$ (Fig. \ref{FConn}a),
  the node $1$
has a strong ability to transmit signals towards the node $5$ (namely the response of this unit is high).
On the contrary, nodes $5,6$ and ...$7$ (the topological hub) have weak performances in signal 
transmission at this frequency. Moreover, one sees that $7$ is a bad sender and a bad receiver: 
in this sense
it is not a hub at this frequency. 
With a frequency $0.57$ (unstable resonances) the effective network has a rather symmetric
structure and basically all units respond to this excitation (however with a different amplitude).
Also, some units present a strong affinity with some others, at a specific frequency. This affinity
is however not completely specific: the unit $3$ ``likes'' the frequency $\omega=0.84$ (Fig. \ref{FConn}c) 
whatever is the unit emitting it (but the best excitation is provided by  unit $7$).  Obviously, one also checks
that for frequencies that do not correspond to resonances (such as $\omega=2.33$ in
Fig.   \ref{FConn}f) the response is essentially inexistent whatever the pair.

Finally, this figure shows that it is basically possible to excite
any unit from any other one in such a way  that this unit (and possibly a few other but \textit{not all} the other units)
have a maximal response. This can be observed in more details in Fig. \ref{FMatMaxRep}. This is a matrix
where the entry $i,j$ (receiver/sender) contains the frequency where the modulus of the susceptibility is maximum (first value) 
and the value of this maximum (second value). Clearly, some units are more ``excitable'' than others (such as $5$).

All these effects are due to a combination of topology and dynamics  and they
cannot be read in the connectivity matrix $\cJ$.

%
%
%

\bigskip

$$
\tiny{
\left[
\baR{ccccccccccccccccc}
(3.1,5.84) & (0.572,8.91) & (0.756,5.97) & (3.14,10.6) & (3.05,6.8) & (3.03,6.3) & (0.724,6.71) & (3.14,7.85) & (3.14,8.7) &\\
 (0.563,9.77) & (0.569,22.2) & (0.795,11) & (0.479,10.7) & (0.577,12.4) & (0.373,5.55) & (0.68,13.4) & (0.569,14.5) & (0.577,6.44) &\\
 (1.35,4.45) & (0.569,8.14) & (1.27,7.13) & (0.719,4.42) & (1.28,8.73) & (1.39,5.05) & (1.41,7.83) & (0.844,7.07) & (1.3,4.5) &\\
 (0.563,15.4) & (0.569,35.3) & (0.795,21.9) & (0.463,20.9) & (0.799,19.8) & (0.816,11.5) & (0.71,19.8) & (0.844,22.1) & (0.482,11.3) &\\
 (1.35,4.6) & (0.569,5.32) & (1.31,7.67) & (0.742,4.04) & (1.28,10.5) & (1.39,5.22) & (1.26,8.3) & (0.844,6.88) & (1.3,5.53) &\\
 (0.36,21.6) & (0.127,28.2) & (0.385,23.5) & (0.437,27.8) & (0.138,17.6) & (0.364,17.6) & (1.26,15.5) & (0.569,19.1) & (0.129,15) &\\
 (0.141,6) & (0.569,10.1) & (1.31,7.58) & (0.437,8.37) & (1.28,9.33) & (1.39,4.53) & (1.26,9.09) & (0.854,6.79) & (1.3,5.23) &\\
 (0.563,6) & (0.569,14.1) & (0.795,7.86) & (0.479,8.9) & (0.563,7.53) & (0.399,4.37) & (0.71,7.91) & (0.569,10) & (0.5,4.61) &\\
 (3.13,19.4) & (3.12,14.9) & (3.02,16.9) & (3.14,32) & (3.04,21.1) & (3.03,20.2) & (3.02,13.4) & (3.14,26.5) & (3.14,25.8) &
\eaR
\right]
}
$$

\vspace{0.5cm}

\bc
\footnotesize{\label{FMatMaxRep} Matrix
where the entry $i,j$ (receiver/sender) contains the frequency where the modulus of the susceptibility is maximum (first value)
and the value of this maximum (second value) }
\ec

\bigskip

[Extrapolating further the possibilities suggested by these figures, one may imagine to apply at node $3$ a superposition
of signals, with amplitude modulation, but with a different carrier frequency (e.g. $\omega_1=0.125$ and $\omega_2=2.33$), such that 
$5$ and $8$ respond simultaneously to their own resonance frequency. However, this operation requires to be strictly
in the linear response regime (small $\epsilon$).]

\ssu{Effect of $\epsilon$.}\label{Ssusvseps}

\begin{figure}[ht!]
\includegraphics[height=8cm,width=12cm,clip=false]{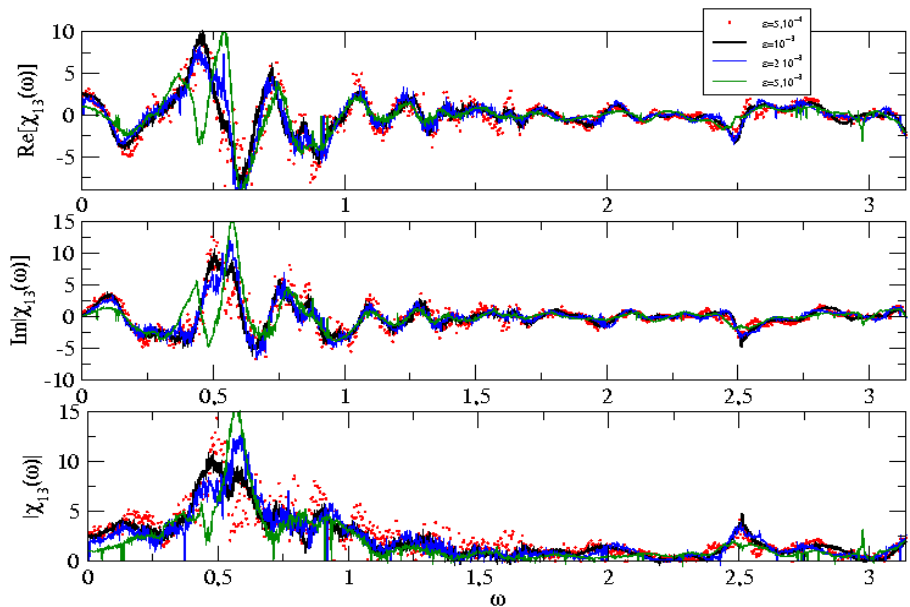}
\caption{\label{FSusvspe}  Effect of increasing $\epsilon$ for the susceptibility $\hchi_{13}$ for $\epsilon
\in [5 \times 10^{-4};5 \times 10^{-3}]$.}
\end{figure}

A linear response theory assumes that $\epsilon$ is small enough, so that $\hchi_{ij}$, corresponding to the first order
term in the $\epsilon$ expansion of $<\delta \bu>$, is independent of $\epsilon$. Henceforth, to check that the value $\epsilon=10^{-3}$
chosen in our simulations is sufficiently small, we have to verify that multiplying or dividing $\epsilon$ by some
(small factor) does not change the susceptibility. Actually, later on we will also be interested in larger values of
$\epsilon$ where the nonlinear terms in the $\epsilon$ expansion are non negligible (see section \ref{SNL}). Although this regime brings 
the system out of the linear response setting, it provides interesting stability properties for amplitude modulation.
However, it is well known that nonlinearities change the resonance structure.
Consequently, we have investigated the influence of increasing $\epsilon$ on the susceptibilities. Some
examples are depicted in Fig. \ref{FSusvspe} for the pair $3 \to 1$
 
%
%
%

One remarks that the susceptibility is stable in the range $[5 \times 10^{-4};2.\times 10^{-3}]$
(note however that the signal is more noisy when $\epsilon$ is weaker, explaining the larger
fluctuations for $\epsilon=5 \times 10^{-4}$). Increasing  $\epsilon$ further leads to distortions
in the resonance curve (see section \ref{SNL}).

\ssu{Amplitude modulation.}\label{Smodamp}

The existence of strong amplitude specific resonances  opens up the possibility for transmitting a signal
carrying information from a node to a target node, in such a way  that, with a suitable filtering of the chaotic background,
 the initial
signal can be recovered. For this, one may use amplitude modulation where the characteristic time scale
for the modulation is sufficiently long. More precisely, one  performs the average   (\ref{susth})
with a sliding time window whose width is sufficiently large to have small fluctuations but remains sufficiently small
compared to the characteristic time for the modulation. 

To illustrate this point we have superimposed a signal with periodic amplitude modulation,
$\xi(t)=\epsilon cos(\omega_M t)sin(\omega_0 t)$
where $\omega_0$ is a resonance frequency and $\omega_M$ the  frequency of the amplitude  modulation. 
Note that, in the case $\epsilon=10^{-3}$ the variations of the signal amplitude stay within
the limits where linear response theory applies. However, for such a weak signal amplitude, the signal/noise ratio
is very large and we had to perform the average (\ref{susth}) over a time windows of width $T=10^6$. This imposes
strong constraints on the modulation frequency, which has to be smaller than $\frac{\pi}{T}$ to have a correct
sampling of the signal. The simulations have been done with $\omega_M=\frac{2\pi}{2.1123 T} \sim 2.97. 10^{-6}$.
The (arbitrary) non integer factor $2.1123$ has been introduced to avoid commensurability between the frequency
corresponding to the sliding window ($\frac{2\pi}{T}$) and $\omega_M$.\\

\begin{figure}[!ht!]
\includegraphics[height=8cm,width=12cm,clip=true,angle=0]{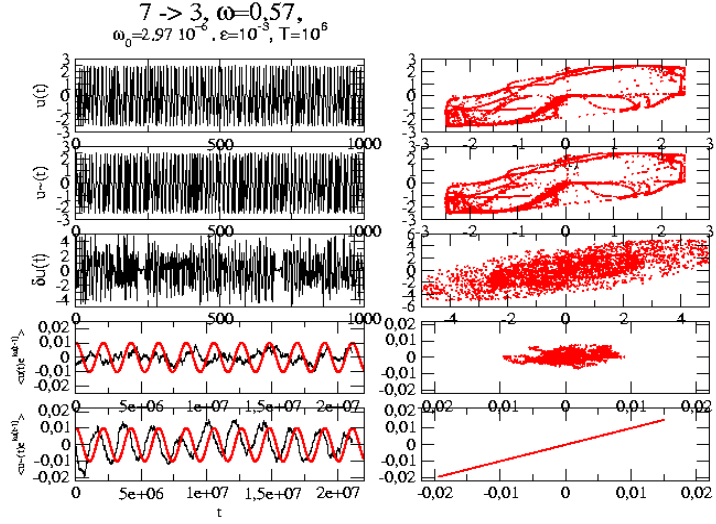}
\caption{\label{FModamp} Excitation of $3$ by $7$ with a signal $\xi(t)=\epsilon cos(\omega_M t)sin(\omega_0 t)$ with
a fundamental frequency $\omega_0=0.7$ and a modulation frequency $\omega_M \sim 2.97. 10^{-6}$. 
The column on the left represents, the unperturbed local field  $u_3(t)$, the perturbed local field $\tu_3(t)$,
the difference between the two local fields $\delta u(t)=u_3(t)-\tu_3(t)$, the time average (\ref{susth}) with a time window 
$T=10^6$ for $u_3(t)$ (4th row) and $\tu_3(t)$ (5th row), respectively. The signal is represented in red, with a
magnification by a factor 10, in order to see it on the graph. The second column represents the same quantities with representation
$u(t+1)$ vs $u(t)$ giving a two dimensional projection of the dynamics in the phase space. .}
\end{figure}

We have first considered (Fig. \ref{FModamp}) the case $7 \to 3$ with a resonance frequency $\omega=0.57$ (see Fig.\ref{FResonances}a).
One clearly sees that the signal can be recovered by the averaging procedure (\ref{susth}). Note also
that one gets an effective amplification by a factor $\sim 20$ in agreement with the resonance curve \ref{FResonances}a.
Consequently, and contrary to  conventional intuition about chaotic systems, a weak signal superimposed
upon a chaotic background can be recovered provided one performs a suitable average over the chaotic dynamics.
In some sense, this idea is already contained in Boltzmann's work
where macroscopic observable values are obtained  by averaging over the microscopic molecular chaos. 

%
%
%

%

\begin{figure}[!ht!]
\includegraphics[height=8cm,width=12cm,clip=true]{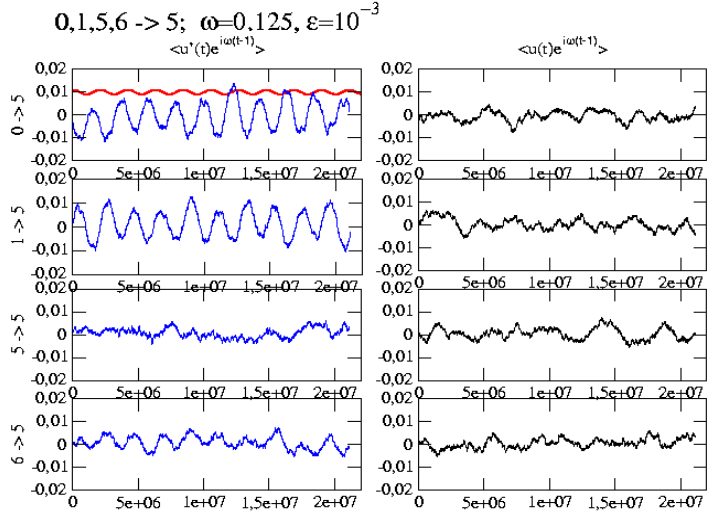}
\caption{\label{FReponse_5_Exc_all_om0.125}  First column. Response of unit $5$ respectively to an excitation of the units $0$,$1$,$5$ and $6$,
 with
amplitude modulation. The fundamental frequency is $\omega_0=0.125$ and the modulation frequency is $\omega_M=2.97 10^{-6}$.
The second column represents the same average performed on the unperturbed dynamics. Note that the average corresponding to each pair have
been performed for different initial conditions.}
\end{figure}

We have then investigated the possibility of sending  a signal from a unit to some target by suitably selecting the frequency.
In section \ref{Ssus} we have given the example of exciting $5$ with a frequency $0.125$ or $1.26$. In the first case, it is expected
that a signal emitted from $0$ or $1$ will be correctly received by $5$ while the same signal emitted from $5$ or $6$ 
will not be distinguished
from the chaotic background. This is verified in Fig. \ref{FReponse_5_Exc_all_om0.125}. The first column represents the response
after performing the average (\ref{susth}) on the perturbed system $\tu$. The right column represents the same average
performed on the unperturbed system (without signal) $u$. 
It is clear that the signal emitted by the units
$0,1$ is correctly recovered by unit $5$ (with however some distortions) while there  is no clear difference
between the perturbed and unperturbed cases when the units $5,6$ are emitting the same signal.
Note that the average corresponding to each pair have
been performed for different initial conditions. This explains why the figures in the right column are different.
%
%
%

The  case $\omega=1.26$ is presented in the Fig. \ref{FReponse_5_Exc_all_om1.26}

%
%
%

\ssu{nonlinear regime.}\label{SNL}

The main drawback of the previous examples is the weakness of the signal. In order to have an efficient elimination 
of the chaotic background one needs to average over a long time window, limiting de facto the modulation frequencies,
 and, even so, the decoded signal is not completely satisfactory. It is then reasonable to increase the amplitude 
of the signal. But first one has then to check that the resulting dynamics remains chaotic. Indeed,  too large a signal
will irretrievably ``kill'' the background. Even when $\epsilon$ is weak enough so that one may still consider the signal as a perturbation,
increasing the signal/noise ratio can drive the system outside the linear response regime. Indeed, as suggested in section \ref{Ssusvseps}
the susceptibility curves are modified when $\epsilon$ is larger than $5. 10^{-3}$. In this section we investigate this effect
more carefully.

\begin{figure}[ht!]
\includegraphics[height=5cm,width=14cm,clip=false]{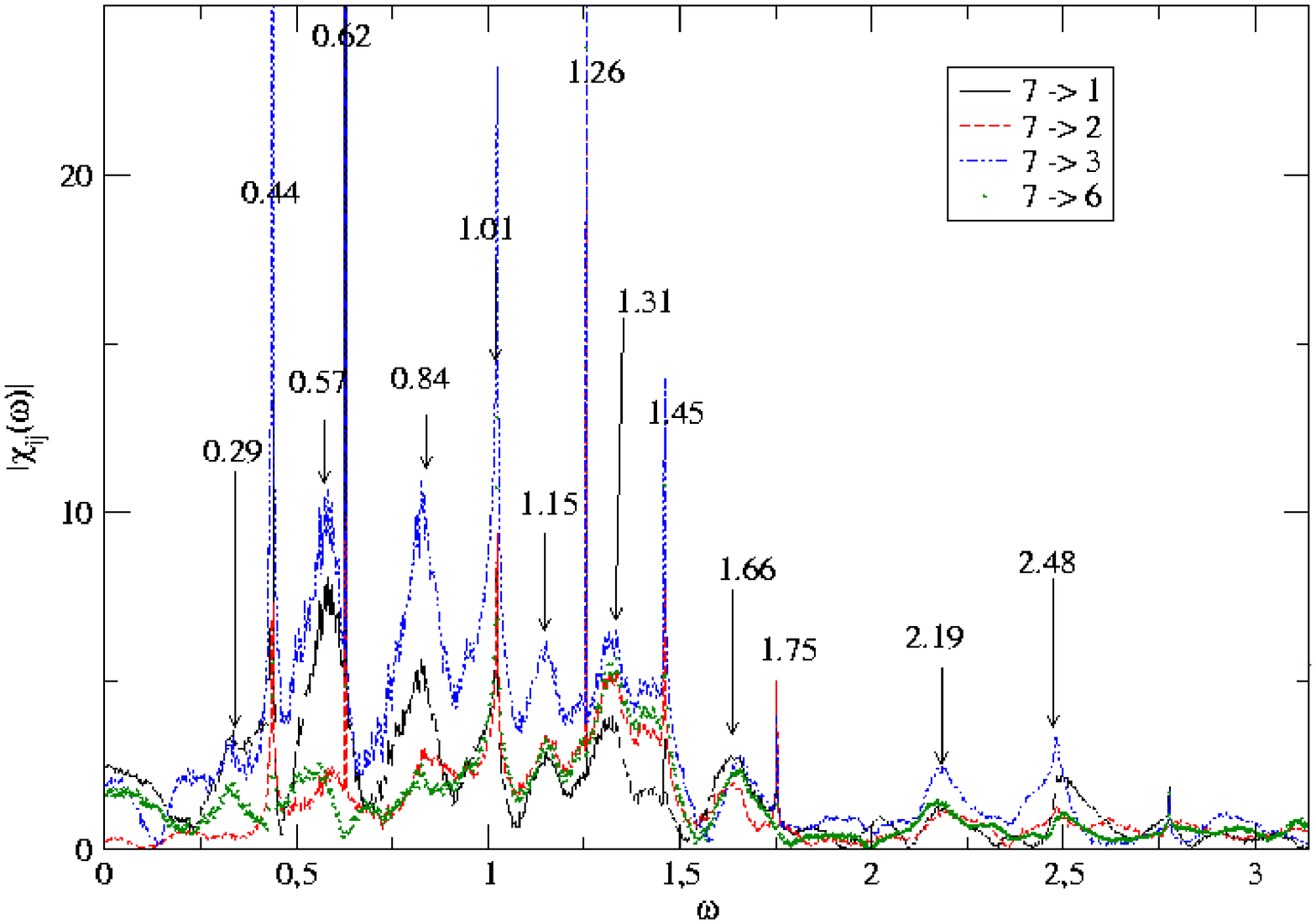}
\vspace{0.2cm}
\includegraphics[height=5cm,width=14cm,clip=false]{Susceptibilite_Exc0_pe1E-2}
\vspace{0.2cm}
\includegraphics[height=5cm,width=14cm,clip=false]{Susceptibilite_Rep5_pe1E-2}
\caption{\label{FSuseps0.01}. Resonance curves in the same cases as in Fig. \ref{FResonances}a,b,c, for $\epsilon=10^{-2}$.}
\end{figure}

\begin{figure}[!ht!]
\includegraphics[height=8cm,width=12cm,clip=false]{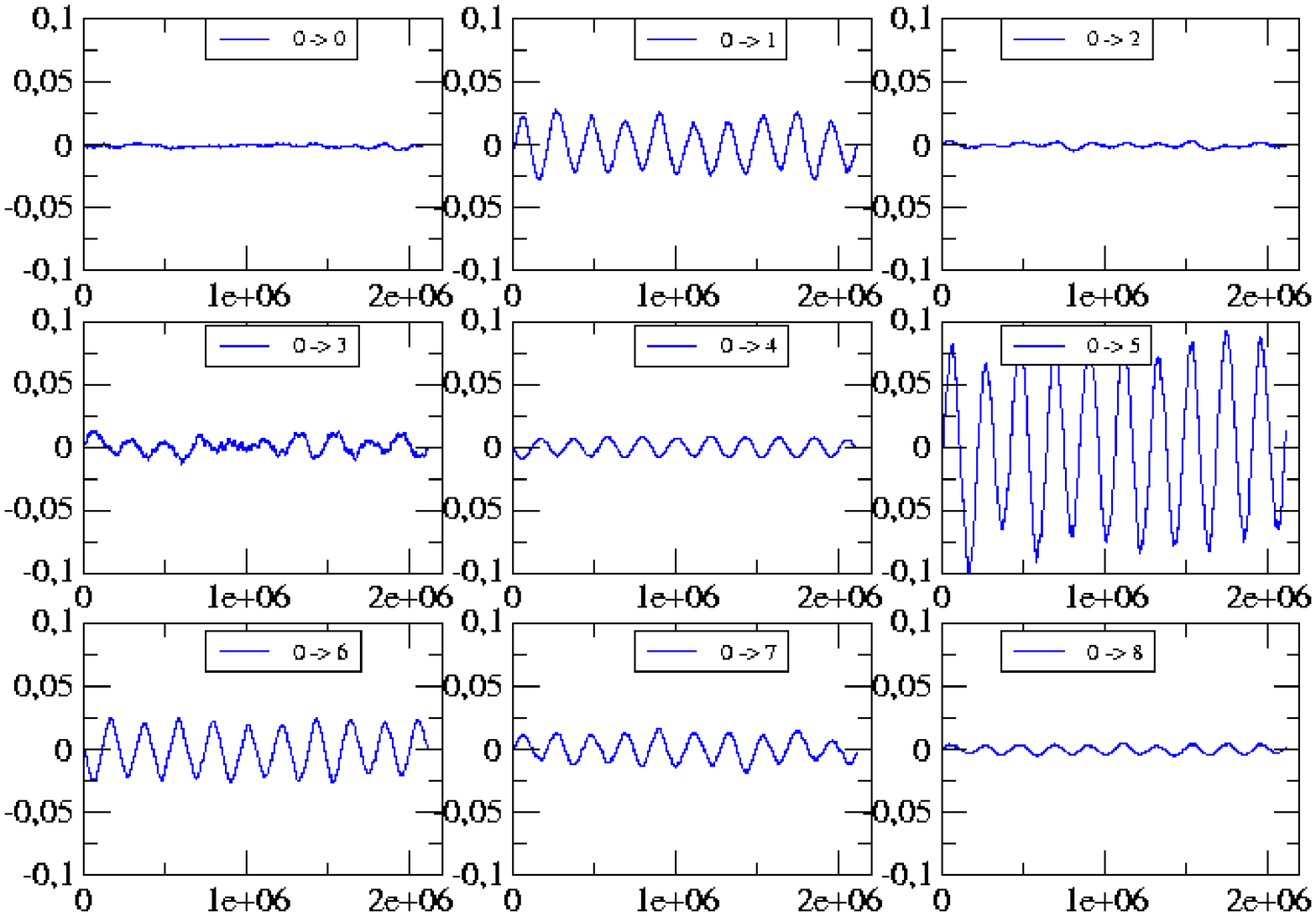}
\caption{\label{FCompare_mod_amp_exc0}  Response of all units  to an excitation of the units $0$ with
amplitude modulation and $\epsilon=0.01$. 
The fundamental frequency is $\omega=0.32$ and the modulation frequency is $\omega_0=2.97 10^{-5}$.}
\end{figure}

First we note that an explicit formula for nonlinear corrections have been worked out by Ruelle in \cite{NLRuelle}. However,
it is hardly tractable, even in the case of model (\ref{DNu}) where the linear response has a simple form. We used then the numerical
computation (\ref{susth}) for $\epsilon=10^{-2}$. In Fig. \ref{FSuseps0.01}a,b,c we have represented
the susceptibility curves for the same cases as in Fig. \ref{FResonances}a,b,c, section \ref{Ssus}. One observes 
sharper resonance peaks. This corresponds to having poles approaching the real axis when increasing the strength of the periodic forcing.
Certainly, for sufficiently large $\epsilon$ and for specific resonant frequencies, one expects the dynamics to ``lock'' on a periodic 
orbit with the effect of ``killing'' the chaotic regime. This is however not yet the situation for the value of $\epsilon$ investigated here,
as verified here (see e.g. Fig. \ref{FCompare_mod_amp_exc0}).
%
%
%

%
%
%

%
%
%

As an example, we have excited the unit $0$ with a frequency $\omega=0.32$, corresponding to a sharp resonance
and with an amplitude modulation frequency $\omega_0=2.97 10^{-5}$ (note that this frequency is ten times higher
than the previous one. We were then able to shrink the sliding window by a factor $10$).
The response of the $9$ units is plotted in Fig. \ref{FCompare_mod_amp_exc0}. The signal/noise ratio is substantially better.
We also observe that several units respond, but the more accurate response corresponds to the unit $5$. 

\begin{figure}[!ht!]
\includegraphics[height=6cm,width=12cm,clip=false]{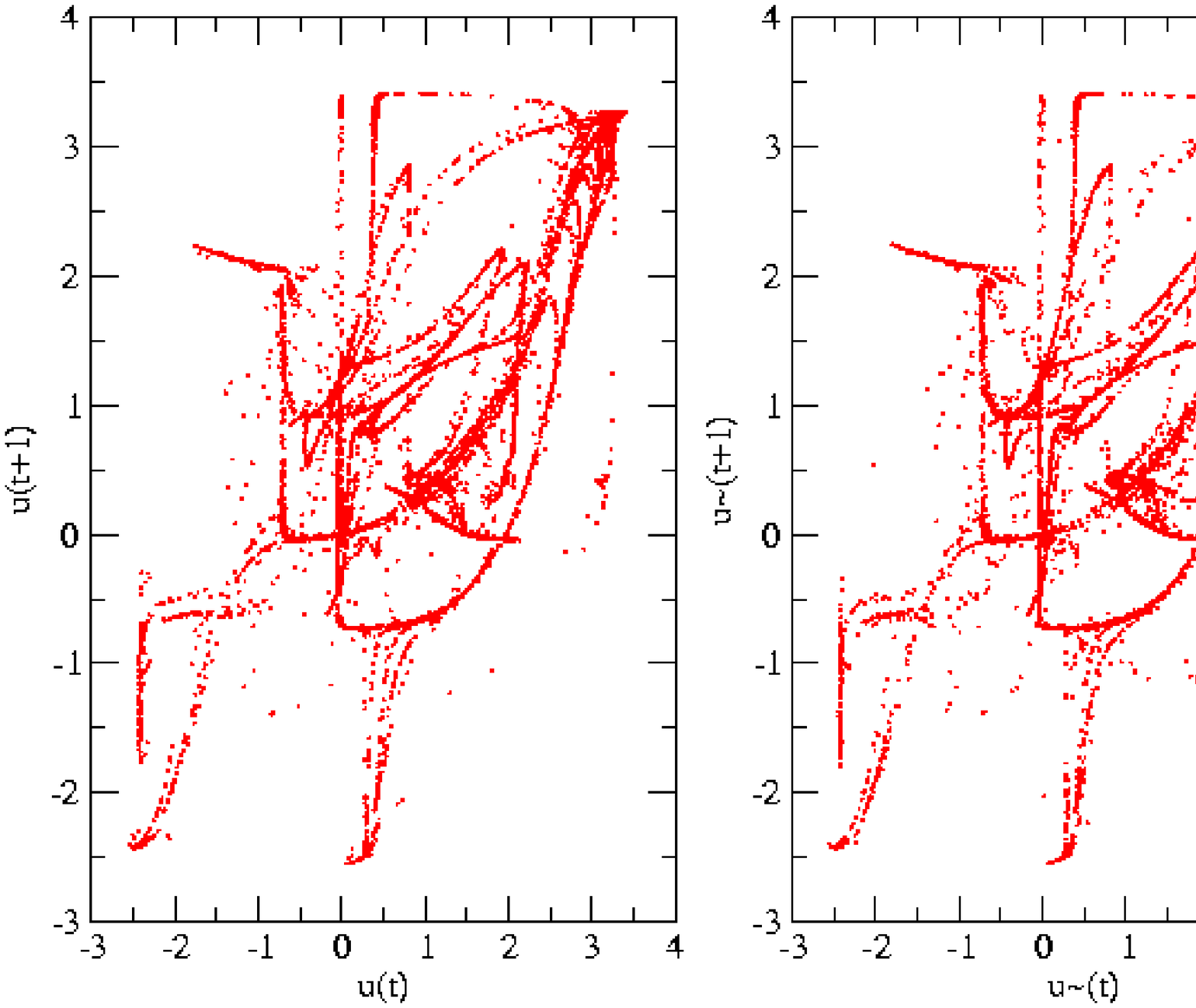}
\caption{\label{FCompare_attracteurs_50_omega0.32_pe0.01} Projection  of the trajectories
$u_5(t)$ (unperturbed system) and $\tu_5(t)$ (perturbed system with amplitude modulation,
$\epsilon=0.01$, $\omega=0.32$ and  modulation frequency is $\omega_0=2.97 10^{-5}$).}
\end{figure}

How does
 the perturbed evolution of this unit look like? The unperturbed and perturbed trajectories are plotted in Fig. \ref{FCompare_attracteurs_50_omega0.32_pe0.01}.
One does not see any difference. In particular the perturbed dynamics is still chaotic.

Assume now that there is a user at node $5$, observing the dynamics. Without filtering, he does not notice
any difference between the system with and without signal. But, if he knows the carrier frequency, he is able to recover
a signal emitted from the unit $0$ out of the chaotic background. This suggests a way to encode hidden information
in a chaotic signal. There is in fact a little bit more. The same user located at $2,3$ or $8$ will not be able to recover
a sufficiently good signal. In this sense, the nonlinearity allows us to send the signal to specific targets by a suitable
choice of the frequency modulation.

\pagebreak

\su{Discussion.}\label{Sdicuss}

In this paper, we have exhibited some  non-intuitive
aspects of networks with nonlinear relays and chaotic dynamics.  
These effects were predicted on the basis of general theoretical arguments,
analyzing the linear response of such systems to an excitation by
a signal injected at some place in the network. We have in particular
argued that saturation effects in the nonlinear transfer function
of the node induces the presence of resonances, the stable ones,
that are not present in the correlation functions. These resonances, corresponding
to the response of the system to out-of-equilibrium perturbations, can be
used to produce unexpected results  in chaotic systems, such as the transmission 
of a signal with amplitude modulation from a node to some target. Though the
signal is basically weak, it can be recovered by a suitable averaging procedure,
in spite of chaos. Moreover, thanks to chaos, this recovery can only be performed
if the receiving user knows the carrier frequency and if he is located at the right node.
Note that this transmission is robust with respect to noise, as we checked.

Furthermore, it would be most interesting to realise an implementation of this scheme on ``real''
experimental chaotic networks. For example, the frequency dependent averaging of eq. (\ref{susth})
could be implemented with a ``lock-in'' amplifier \cite{Libbrecht}.  
  
We have presented an example supporting these conjectures. 
We have briefly discussed in section \ref{SignProp} the genericity of this example.
But what about the genericity of the \textit{model} itself?
 As discussed in the introduction this model contains some essential
 features such as the competition of excitation/inhibition, the asymmetry of the interactions,
and the saturation of the transfer functions that are basically 
present in biological networks or in some communication networks.
These features generically produces chaos in systems like (\ref{DNu}), provided that the
nonlinearity is sufficiently large (see  \cite{DNnet} for a recent review).
However, one does not necessarily have chaos
in ``real'' networks. It might indeed well be that biological networks, for example,
are often closer
to intermittency than to chaos and closer to bifurcation points than to structurally stable hyperbolic
systems (see \cite{Bak,DNnet}). In this case, the application of the methods developed here may lead to
fundamental questions such as: do we still have a linear response theory in this case ?
As discussed in the paper, when one approaches  a structurally unstable point (bifurcation point)
the  susceptibility may diverge, as it does in physical systems  at a second order phase transition.
Note however, that the \textit{the way the susceptibility diverges} is already a crucial information.
 Actually, in systems undergoing a second order phase transition, the divergence
occurs if one takes the thermodynamic limit, but physical systems are finite. In the same way,
the divergence of our susceptibility may occur when one takes the infinite time limit, but real networks
are investigated on finite times. 
This question deserves therefore further investigations.  \\

Nevertheless, the simple fact that we have been able to produce one example 
of the effects theoretically predicted for such networks, obliges, in our opinion, the community
to review some a priori. Though it is helpful to investigate the topological properties
of complex networks (small world, scale free graphs and so on) it is in no way sufficient
for characterizing the  ability of transmission of such networks with \textit{active} nodes.
Moreover, the dynamics of signal propagation \textit{is not} a superposition of the
graph properties and of the local input/output dynamics. There is a  complex nonlinear
feedback between the two (as revealed for example in eq. (\ref{chiij})) and one has to study
it as a whole. This may require the development of new tools such as the linear response
theory presented here. 
The development of such tools opens new perspectives and suggest new applications for 
communication networks, as well as a better understanding of biological networks.

\ed